\documentclass[%
reprint,
superscriptaddress,
nofootinbib,
amsmath,
amssymb,
aps,
prd,
floatfix,
showkeys,onecolumn]{revtex4-2}

\usepackage{graphicx}% Include figure files
\usepackage{dcolumn}% Align table columns on decimal point
\usepackage{bm}% bold math
\usepackage[normalem]{ulem}
\usepackage{fancyhdr}
\usepackage{graphicx,amsfonts,amssymb,amsbsy}
\usepackage{amsmath,amsthm,latexsym}
\usepackage[utf8]{inputenc}
\usepackage{natbib}
\usepackage[colorlinks]{hyperref}
\usepackage{booktabs}
\usepackage{xcolor}
\usepackage{xspace}
\usepackage{soul}
\usepackage{comment}
\usepackage{caption}
\hypersetup{
    citecolor=blue,  % Color de las referencias bibliográficas
    linkcolor=red,  % Color de los links internos (ecuaciones, figuras, secciones)
    urlcolor=blue    % Color de los enlaces a sitios web
}

\begin{document}

\title{Higher-order generalized uncertainty principle corrections to Casimir-supported traversable wormholes}

\author{Jureeporn Yuennan}
\email{jureeporn\_yue@nstru.ac.th}
\affiliation{Faculty of Science and Technology, Nakhon Si Thammarat Rajabhat University, \\Nakhon Si Thammarat, 80280, Thailand}

\author{Allah Ditta}
\email{mradshahid01@gmail.com}
\affiliation{School of Science, Walailak University, Nakhon Si Thammarat, 80160, Thailand}
\affiliation{Center of Excellence in High Energy Physics, Faculty of Science, Chulalongkorn University, Phayathai Road, Pathumwan, Bangkok 10330, Thailand}

\author{Thammarong Eadkhong}
\email{thammarong.ea@mail.wu.ac.th}
\affiliation{School of Science, Walailak University, \\Nakhon Si Thammarat, 80160, Thailand}

\author{Kazuharu Bamba}
\email{bamba@sss.fukushima-u.ac.jp}
\affiliation{Faculty of Symbiotic Systems Science, Fukushima University, Fukushima 960-1296, Japan}

\author{Phongpichit Channuie}
\email{phongpichit.ch@mail.wu.ac.th}
\affiliation{School of Science \& College of Graduate Studies, Walailak University, \\Nakhon Si Thammarat, 80160, Thailand}

\begin{abstract}
We investigate traversable wormholes supported by Casimir vacuum energy with second-order generalized uncertainty principle (GUP) corrections. For two representative GUP models, we derive higher-order corrections to the Casimir energy and construct exact wormhole solutions in general relativity. The resulting geometries satisfy the throat, flare-out, and asymptotic-flatness conditions. Higher-order corrections modify the wormhole mass and reduce the exotic matter required, although the null and weak energy conditions remain violated near the throat. We further analyze weak gravitational lensing and gravitational-wave echoes, finding model-dependent signatures that may distinguish different GUP realizations. By relating the dimensionless parameter used in the solutions to the conventional phenomenological GUP parameter, we show that its physical interpretation depends strongly on the throat radius. Most current experimental bounds favor near-Planckian throats for appreciable GUP corrections, while macroscopic throats generally require much weaker corrections. These results provide a framework for confronting GUP-corrected Casimir wormholes with laboratory constraints and astrophysical observations.
\end{abstract}

\maketitle

%------------------------------------------------------------------
\section{Introduction}
\label{sec:intro}
%-------------------------------------------------------------------

One of the central challenges in constructing traversable wormholes \cite{Morris:1988cz} is identifying a physically viable source of exotic matter capable of satisfying the flare-out condition while violating the classical energy conditions. Two complementary approaches have been extensively explored in the literature. The first exploits modified theories of gravity, in which higher-curvature terms or additional gravitational degrees of freedom effectively mimic exotic matter. Representative examples include higher-order curvature theories, Einstein-Gauss-Bonnet gravity, teleparallel gravity, $f(R)$ gravity, Horava--Lifshitz gravity, massive gravity, and scalar-tensor theories, among many others~\cite{Giribet:2019dmg,Ovgun:2017dik,KordZangeneh:2015dks,Mehdizadeh:2012zz,Dehghani:2011fa,Garcia-Compean:2020aaa,Kamma:2021wam,Mishra:2021ato,Korolev:2020ohi,Sharif:2020xxj,Singh:2020rai}. The second approach remains within Einstein gravity and introduces suitable exotic matter sources, such as vacuum energy, cosmological constant, phantom or quintessence scalar fields, nonlinear gauge fields, Chaplygin gas fluids~\cite{Santos:2021jjs,Garattini:2019ivd,Aounallah:2020rlf,Anand:2020wlk,Lobo:2005vc}, other beyond Einstein gravities \cite{Dehghani:2009zza,Ghoroku:1992tz,MenaMarugan:1991ea,Hochberg:1990is,Bellorin:2014qca,Botta-Cantcheff:2009ffi,Paul:2018ppy,Jusufi:2017drg}, $f(R)$ gravity \cite{Eid:2020hvf,Shamir:2020uzy,Fayyaz:2020jzh,Tangphati:2020mir,Godani:2019kgy,Samanta:2019tjb,Godani:2018blx,Sharif:2018jdj,Saeidi:2011zz,Bronnikov:2010tt,Lobo:2009ip} and scalar-tensor theories \cite{Papantonopoulos:2019ugr,Franciolini:2018aad,Mironov:2018uou,Mironov:2018pjk,Evseev:2017jek,Rubakov:2016zah,Kolevatov:2016ppi,Bhattacharya:2009rt,Lu:2003bg,He:2002bb,Nandi:1997en,Agnese:1995kd,Xiao:1991nv}. Additionally, there are wide classes of wormholes classified by their spacetime geometry, stability, and the underlying physical frameworks that support them, see e.g.,~\cite{Ibadov:2020btp,Antoniou:2019awm,Mehdizadeh:2015jra,Kanti:2011yv,Kanti:2011jz,Mazharimousavi:2010bm,Maeda:2008nz,Bhawal:1992sz}.

Among the various candidates proposed to provide the exotic matter required for traversable wormholes, the Casimir effect occupies a distinguished position because it represents one of the few experimentally verified manifestations of quantum vacuum fluctuations. The Casimir vacuum naturally possesses a negative energy density arising from quantum field theory in the presence of boundaries, making it a physically well-motivated alternative to phenomenological exotic matter models. In particular, Garattini~\cite{Garattini:2019ivd} demonstrated that Casimir vacuum energy alone is sufficient to support asymptotically flat traversable wormholes, thereby establishing a direct connection between experimentally confirmed quantum vacuum phenomena and wormhole physics. From the perspective of quantum gravity, however, the Casimir vacuum is expected to receive corrections at sufficiently short distances owing to the existence of a fundamental minimal length. Such modifications arise naturally within the framework of the generalized uncertainty principle (GUP), where deformed canonical commutation relations lead to corrections to the standard Casimir energy. Since the geometry of Casimir-supported wormholes is determined directly by the vacuum energy density, these quantum-gravity corrections are expected to modify both the spacetime structure and its physical properties. Consequently, GUP-corrected Casimir wormholes provide a natural theoretical framework for investigating how minimal-length effects propagate from microscopic quantum vacuum fluctuations to macroscopic gravitational configurations, see a first-order GUP contribution to WH solutions~\cite{Jusufi:2020rpw}.

A first step in this direction was carried out by considering leading-order corrections induced by the generalized uncertainty principle (GUP), where it was shown that minimal-length effects modify the Casimir vacuum energy and consequently alter the corresponding wormhole solutions. Although these first-order analyses established the qualitative influence of GUP on Casimir wormholes, they leave open an important question concerning the role of higher-order quantum-gravity corrections. Since the GUP is fundamentally a perturbative deformation of the canonical commutation relations, there is no {\it a priori} reason to expect the leading-order contribution to capture the complete physical behavior. Higher-order terms may generate non-negligible corrections to the spacetime geometry, gravitational mass, and observable signatures, while simultaneously revealing the underlying perturbative structure of the quantum vacuum.

Motivated by these considerations, in the present work we extend the GUP-corrected Casimir energy beyond the leading-order approximation by deriving analytical expressions up to second order in the minimal-length parameter for two representative GUP models. We show that the resulting Casimir energy naturally admits a systematic perturbative expansion in even powers of the dimensionless parameter $\hbar\sqrt{\beta}/a$, with coefficients determined by the underlying realization of the generalized uncertainty principle. Employing these corrected energy densities as the matter source in the Einstein field equations, we construct exact traversable wormhole solutions and investigate the effects of higher-order quantum-gravity corrections on their geometry, energy conditions, ADM mass, and exotic matter content. We further examine several theoretical observational signatures, including embedding diagrams, weak gravitational lensing, and gravitational-wave echoes, in order to assess whether different realizations of the GUP can lead to distinguishable astrophysical predictions.

The remainder of this paper is organized as follows. In Sec.~\ref{sec2} we revisit the Casimir effect in the presence of the generalized uncertainty principle and derive analytical second-order corrections for two representative GUP models. In Sec.~\ref{sec3} we employ these corrected Casimir energy densities to construct traversable wormhole solutions with different redshift functions. Section~\ref{sec4} evaluates the corresponding ADM mass, while Secs.~\ref{sec5} and \ref{sec6} examine the energy conditions and quantify the amount of exotic matter required to sustain the wormhole geometries. In Sec.~\ref{sec:obs} we investigate several theoretical observational signatures, including embedding diagrams and weak gravitational lensing, whereas Sec.~\ref{sec:echoes} discusses gravitational-wave echoes. Sec.~\ref{sec:GUPbounds} connects the range of the GUP parameter adopted in our analysis with existing phenomenological bounds and clarifies the relation between the dimensionless parameter, the dimensional GUP parameter, and the associated minimal-length scale; it further surveys independent experimental and astrophysical bounds on the GUP parameter and derives the crossover throat radius below which each bound still permits order-unity GUP-Casimir corrections. This comparison establishes the regime in which our perturbative treatment remains consistent with current constraints and identifies the parameter range relevant to the wormhole solutions. Finally, Sec.~\ref{sec9} summarizes our main results and outlines possible directions for future research. Throughout this work, we adopt geometrized units with $G=c=1$.

%%%%%%%%%%%%%%%%%%%%%%%%
\section{Casimir effect under higher-order GUP}\label{sec2}
%%%%%%%%%%%%%%%%%%%%%%
The Casimir effect is a remarkable quantum phenomenon that arises from modifications of the vacuum fluctuations of the electromagnetic field in the region between two neutral, perfectly conducting, parallel plates. From a quantum-field-theoretic perspective, this effect originates from the change in the zero-point energy caused by the presence of boundaries. As a consequence, a measurable force develops between the plates, leading to an attractive interaction. In the pioneering work of Ref.~\cite{Casimir:1948dh}, the vacuum energy density (VED) associated with two parallel conducting plates was calculated, yielding the energy per unit area
\begin{eqnarray}
\mathcal{E}_{\rm \small{VED}}=-\frac{\pi^{2}}{720}\frac{\hbar}{a^{3}},
\label{eq:2}
\end{eqnarray}
where $a$ denotes the separation between the plates measured along the $z$-direction, which is orthogonal to the plate surfaces. Differentiating the energy with respect to the plate separation gives the corresponding Casimir pressure,
\begin{eqnarray}
{\cal P}_{\rm \small{VED}}\equiv -\frac{d\mathcal{E}_{0}}{da}=-\frac{\pi^{2}}{240}\frac{\hbar}{a^{4}},
\end{eqnarray}
where the negative sign indicates that the force is attractive. At extremely short distances, the structure of spacetime is expected to be influenced by quantum-gravitational effects. Many approaches to quantum gravity predict the existence of a fundamental minimal length, typically of the order of the Planck scale, which limits the achievable spatial resolution. Such a minimal length also implies the existence of an upper bound on the energy that particles can attain, usually associated with the Planck energy. Consequently, the standard expressions for linear momentum and the canonical commutation relations are expected to receive corrections, leading to modified dispersion relations. Examples of these modifications appear in frameworks such as gravity's rainbow~\cite{Magueijo:2002xx}, with a variety of cosmological~\cite{Chatrabhuti:2015mws,Channuie:2019kus,Hendi:2016tiy} and astrophysical applications~\cite{Hendi:2016hbe,Feng:2017gms,Hendi:2018sbe,Panahiyan:2018fpb,Dehghani:2018qvn}. Within these scenarios, the minimal length naturally manifests itself as a nonzero lower bound on position uncertainty, namely $\Delta x_{0}>0$.

A well-known example arises in string theory, where distances smaller than the characteristic string length cannot be resolved. This observation leads to corrections to the conventional Heisenberg uncertainty principle. In one spatial dimension, the presence of a minimal length can be incorporated through a generalized uncertainty principle (GUP) of the form
\begin{eqnarray}
\Delta x \Delta p
\geq
\frac{\hbar}{2}
\left[
1+\beta(\Delta p)^2+\gamma
\right],
\,\,
\beta,\gamma>0,
\label{cr}
\end{eqnarray}
which predicts a nonvanishing minimal position uncertainty,
\begin{eqnarray}
\Delta x_{0}=\hbar\sqrt{\beta},
\end{eqnarray}
where $\beta$ characterizes the fundamental length scale. The modified uncertainty relation in Eq.~(\ref{cr}) induces corrections to the canonical Heisenberg algebra, resulting in the deformed commutation relation
\begin{eqnarray}
\left[\hat{x},\hat{p}\right]
=
i\hbar
\left(
1+\beta\hat{p}^{2}
+\cdots
\right).
\label{H}
\end{eqnarray}
An important consequence of this deformation is that exact position eigenstates cease to represent physical states in the usual sense. To overcome this difficulty, one commonly employs the quasi-position representation, in which quantum states are projected onto maximally localized states. Furthermore, the deformed algebra can be extended beyond one dimension. In an $n$-dimensional spatial manifold, a general class of commutation relations compatible with a generalized uncertainty principle can be written as~\cite{Frassino:2011aa}
\begin{eqnarray}
\left[\hat{x}_{i},\hat{p}_{j}\right]
=
i\hbar
\left[
f(\hat{p}^{2})\delta_{ij}
+
g(\hat{p}^{2})\hat{p}_{i}\hat{p}_{j}
\right],
\label{RC}
\end{eqnarray}
where $i,j=1,\ldots,n$. The functions $f(\hat{p}^{2})$ and $g(\hat{p}^{2})$ are constrained by physical requirements such as translational and rotational symmetry and therefore cannot be chosen arbitrarily. As emphasized in Ref.~\cite{Frassino:2011aa}, the explicit structure of the corresponding maximally localized states depends on both the dimensionality of the system and the particular realization of the generalized uncertainty principle. In dimensions greater than one, the generalized uncertainty relations are not unique, and distinct choices of the functions $f(\hat{p}^{2})$ and $g(\hat{p}^{2})$ give rise to different GUP models and, consequently, different families of maximally localized states.

%%%%%%%%%%%%%%%%%%%%
\subsection{Model I}
%%%%%%%%%%%%%%%%%%%%
We consider first the model proposed by Ref.~\cite{Frassino:2011aa}. This model has
the functions $f$ and $g$ given by
\begin{eqnarray}
f\left(p^{2}\right)=1+\beta p^{2},\qquad g\left(p^{2}\right)=0.\label{fg1}
\end{eqnarray}
In this model, we compute the energy per unit surface following Ref.~\cite{Frassino:2011aa}:
\begin{eqnarray}
{\cal E}=\frac{1}{(2 \pi )^2 {\hbar}^2}\Big[-\frac{1}{2!}B_{2}G^{1}(0)-\frac{1}{4!}B_{4}G^{3}(0)-\frac{1}{6!}B_{6}G^{5}(0)-\frac{1}{8!}B_{8}G^{7}(0)+...\Big],
\end{eqnarray}
where
\begin{eqnarray}
G(n)&=&-\frac{\pi}{4 \beta ^3 \left(\frac{\pi ^2 \beta  {\hbar}^2 n^2}{a^2}+1\right)}\Bigg[-\frac{\pi ^3 \beta ^{5/2} {\hbar}^2 n^2}{a^2}+\frac{2 \beta ^{5/2} \left(\pi ^2 {\hbar}^2 n^2\right)}{a^2}\tan ^{-1}\left(\frac{\pi  \sqrt{\beta } {\hbar} n}{a}\right)\nonumber\\&&\qquad\qquad\qquad\qquad\qquad-2 \beta ^{3/2} \tan ^{-1}\left(\frac{\pi  \sqrt{\beta } {\hbar} n}{a}\right)+\frac{2 \pi \beta^2 {\hbar} n}{a}-\pi  \beta ^{3/2}\Bigg]\,,
\end{eqnarray}
with $G^{(i)}(0)$ denote the $i$-th derivative of $G$ with respect to $n$, evaluated at $n=0$. It has been shown in Ref.~\cite{Frassino:2011aa} for the limiting values of the derivatives of $G(n)$ at $n=0$ up to first-order correction ${\cal {O}}(\beta)$. However, the higher-order correction terms, e.g., ${\cal {O}}(\beta^{2}),\,{\cal {O}}(\beta^{3})$, can be simply computed to obtain
\begin{eqnarray}
\lim_{n\to 0} G^{(7)}(n)&=&-\frac{5568 \pi^8 \beta ^2 {\hbar}^7}{a^7}\,,\nonumber\\\lim_{n\to 0} G^{(9)}(n)&=&\frac{446976 \pi ^{10} \beta ^3 {\hbar}^9}{a^9}\,.
\end{eqnarray}
The extension to higher orders is actually quite elegant because the entire structure is controlled by the odd derivatives of $G(n)$ in the Euler-Maclaurin expansion. Using the above expressions, the final result up to the second-order correction term in the minimal uncertainty
parameter $\beta$ introduced in the modified commutation relations of Eq.(\ref{H}) can be obtained:
\begin{eqnarray}
{\cal E}_{I}={\cal E}_{\rm \small{VED}}\Bigg[1+\underbrace{\frac{2 \pi^2}{3}\Bigg(\frac{\hbar \sqrt{\beta}}{a}\Bigg)^{2}}_{1{\rm st-order \,\,correction}}+\underbrace{\frac{29\pi^4}{35}\Bigg(\frac{\hbar \sqrt{\beta}}{a}\Bigg)^{4}}_{2{\rm nd-order \,\,correction}}\Bigg]\,.\label{fmod1}
\end{eqnarray}
The first term in Eq.~(\ref{fmod1}) corresponds to the standard Casimir energy previously defined in Eq.~(\ref{eq:2}), derived in the absence of a cut-off function. The second and third terms provide the corrections arising from the minimal length scale of the theory, which notably yields an attractive contribution. From this, the Casimir pressure between the plates is given by ${\cal P} \equiv -\partial {\cal E}/\partial a$:
\begin{eqnarray}
{\cal P}_{I}={\cal P}_{\rm \small{VED}}\Bigg[1+\underbrace{\frac{10 \pi^2}{9}\Bigg[\frac{\hbar \sqrt{\beta}}{a}\Bigg)^{2}}_{1{\rm st-order \,\,correction}}+\underbrace{\frac{29 \pi^4}{15}\Bigg(\frac{\hbar \sqrt{\beta}}{a}\Bigg)^{4}}_{2{\rm nd-order \,\,correction}}\Bigg]\,.\label{Pmod1}
\end{eqnarray}
It is worth emphasizing that the Casimir energy is characterized by a natural equation of state (EoS), which is recovered by setting $w = 3$. Using the previous relation, the 2nd-order GUP-corrected Casimir energy density can be expressed in the compact form
\begin{eqnarray}
\varrho_{I}=\varrho_{\rm \small{VED}}
\Bigg[1+\underbrace{\frac{10 \pi^2}{9}\Bigg(\frac{\hbar \sqrt{\beta}}{a}\Bigg)^{2}}_{1{\rm st-order \,\,correction}}+\underbrace{\frac{29 \pi^4}{15}\Bigg(\frac{\hbar \sqrt{\beta}}{a}\Bigg)^{4}}_{2{\rm nd-order \,\,correction}}\Bigg],\label{mo1}
\end{eqnarray}
with $\varrho_{\rm \small{VED}}=-\tfrac{\pi^{2}\hbar}{720a^{4}}$. In the limit $\beta \to 0$, the correction term vanishes and the standard Casimir energy density is recovered.

%%%%%%%%%%%%%%%%%%%%%%
\subsection{Model II}
%%%%%%%%%%%%%%%%%%%%%%%
The precise form of these states is determined by the number of dimensions and the specific model under consideration. Here, we adopt another choice of the generic functions $f(\hat{p}^{2})$ and $g(\hat{p}^{2})$ introduced in Ref.~\cite{Frassino:2011aa}:
\begin{eqnarray}
f\left(p^{2}\right)=\frac{\beta p^{2}}{\sqrt{1+2\beta p^{2}}-1},\quad g\left(p^{2}\right)=\beta\,.\label{eq:4}
\end{eqnarray}
In this model, we also compute the energy per unit surface using expressions given in Ref.~\cite{Frassino:2011aa}:
\begin{eqnarray}
{\cal E}=\frac{1}{(2 \pi )^2 {\hbar}^2}\Big[-\frac{1}{2!}B_{2}G^{1}(0)-\frac{1}{4!}B_{4}G^{3}(0)-\frac{1}{6!}B_{6}G^{5}(0)-\frac{1}{8!}B_{8}G^{7}(0)+...\Big],
\end{eqnarray}
where
\begin{eqnarray}
G(n)=-\frac{\left(2 \pi ^4 \sqrt{2} 2^{3 \alpha +\frac{13}{2}} {\hbar}^2 n\right) \sqrt{\frac{\beta {\hbar}^2 n^2}{a^2}}}{\sqrt{\beta } \Big(a^2+2 \pi ^2 \beta {\hbar}^2 n^2\Big)}\left(\sqrt{\frac{a^2+2 \pi ^2 \beta  {\hbar}^2 n^2}{a^2}}+1\right)^{-3 \alpha -7}\,,
\end{eqnarray}
with $\alpha = 1+\sqrt{1+\frac{3}{2}}$. It has been derived in Ref.~\cite{Frassino:2011aa} for the limiting values of the derivatives of $G(n)$ at $n=0$ up to first-order correction ${\cal {O}}(\beta)$. However, the higher-order correction terms, e.g., ${\cal {O}}(\beta^{2}),\,{\cal {O}}(\beta^{3})$, can be simply computed to obtain
\begin{eqnarray}
\lim_{n\to 0} G^{(7)}(n)&=&-\frac{90 \left(93 \sqrt{10}+529\right) \pi ^8 \beta ^2 {\hbar}^7}{a^7}\,,\nonumber\\\lim_{n\to 0} G^{(9)}(n)&=&\frac{1260 \left(1789 \sqrt{10}+8202\right) \pi^{10} \beta ^3 {\hbar}^9}{a^9}\,.
\end{eqnarray}
Using the above expressions, the final result up to the second-order correction term in the minimal uncertainty
parameter $\beta$ introduced in the modified commutation relations of Eq.~(\ref{H}) can be obtained:
\begin{eqnarray}
{\cal E}_{II}={\cal E}_{\rm \small{VED}}\Bigg[1+\underbrace{\left(\frac{28+3 \sqrt{10}}{14}\right)\pi^2 \Bigg(\frac{\hbar \sqrt{\beta}}{a}\Bigg)^{2}}_{1{\rm st-order \,\,correction}}+\underbrace{\left(\frac{1587+279 \sqrt{10}}{224}\right)\pi^4\Bigg(\frac{\hbar \sqrt{\beta}}{a}\Bigg)^{4}}_{2{\rm nd-order \,\,correction}}\Bigg]\,.\label{fmod2}
\end{eqnarray}
The first term in Eq.~(\ref{fmod2}) corresponds to the standard Casimir energy previously defined in Eq.~(\ref{eq:2}), derived in the absence of a cut-off function. The second and third terms provide the corrections arising from the minimal length scale of the theory, which notably yields an attractive contribution. From this, the Casimir pressure between the plates is given by $P\equiv -\partial {\cal E}/\partial a$:
\begin{eqnarray}
P_{II}=P_{\rm \small{VED}}\Bigg[1+\underbrace{\left(\frac{10}{3}+\frac{5 \sqrt{10}}{14}\right)\pi^{2}\Bigg(\frac{\hbar \sqrt{\beta}}{a}\Bigg)^{2}}_{1{\rm st-order \,\,correction}}+\underbrace{\left(\frac{529+93 \sqrt{10}}{32}\right)\pi^4\Bigg(\frac{\hbar \sqrt{\beta}}{a}\Bigg)^{4}}_{2{\rm nd-order \,\,correction}}\Bigg]\,.\label{Pmod1b}
\end{eqnarray}
The Casimir energy is characterized by a natural equation of state (EoS), which is recovered by setting $w = 3$. Therefore, the 2nd-order GUP-corrected Casimir energy density can be expressed in the compact form
\begin{eqnarray}
\varrho_{II}=\varrho_{\rm \small{VED}}
\Bigg[1+\underbrace{\left(\frac{10}{3}+\frac{5 \sqrt{10}}{14}\right)\pi^{2}\Bigg(\frac{\hbar \sqrt{\beta}}{a}\Bigg)^{2}}_{1{\rm st-order \,\,correction}}+\underbrace{\left(\frac{529+93 \sqrt{10}}{32}\right)\pi^4\Bigg(\frac{\hbar \sqrt{\beta}}{a}\Bigg)^{4}}_{2{\rm nd-order \,\,correction}}\Bigg],
\end{eqnarray}
with $\rho_{\rm \small{VED}}=-\tfrac{\pi^{2}\hbar}{720a^{4}}$. In the limit $\beta \to 0$, the correction term vanishes and the standard Casimir energy density is recovered.

%%%%%%%%%%%%%%%%%
\section{GUP-corrected Casimir wormholes}\label{sec3}
%%%%%%%%%%%%%%%%
We consider a static, spherically symmetric traversable wormhole described by the Morris-Thorne metric in Schwarzschild coordinates~\cite{Morris:1988cz},
\begin{equation}
ds^{2}=-e^{2\Phi(r)}dt^{2}+\frac{dr^{2}}{1-\frac{b(r)}{r}}
+r^{2}\left(d\theta^{2}+\sin^{2}\theta, d\phi^{2}\right),
\label{5}
\end{equation}
where $\Phi(r)$ and $b(r)$ denote the redshift and shape functions, respectively. For a traversable wormhole geometry, the redshift function must remain finite throughout the spacetime to prevent the appearance of event horizons. The shape function characterizes the spatial structure of the wormhole and satisfies the throat condition $b(r_{0})=r_{0}$, where $r_{0}$ represents the throat radius.
Furthermore, the wormhole geometry must fulfill the flaring-out condition~\cite{Morris:1988cz},
\begin{equation}
\frac{b(r)-r b'(r)}{b^{2}(r)}>0,
\end{equation}
with the additional requirement that $b'(r_{0})<1$ at the throat. These conditions ensure that the wormhole remains open and traversable.
Using the metric~\eqref{5}, the Einstein field equations yield the following relations for the energy density and pressure components,
\begin{widetext}
\begin{eqnarray}
\rho (r) &=&\frac{1}{8\pi r^{2}} b^{\prime }(r), \label{rho}\\
\mathcal{P}_{r}(r) &=&\frac{1}{8\pi}\left[ 2\left( 1-\frac{b(r)}{r}\right)
\frac{\Phi ^{\prime }}{r}-\frac{b(r)}{r^{3}}\right]
,\label{Pr}  \\
\mathcal{P}_t(r) &=&\frac{1}{8\pi }\left[ 1-\frac{b(r)}{r}\right] \Big[\Phi
^{\prime \prime }+(\Phi ^{\prime })^{2}-\frac{b^{\prime }r-b}{2r(r-b)}\Phi
^{\prime } -\frac{b^{\prime }r-b}{2r^{2}(r-b)}+\frac{\Phi ^{\prime }}{r%
}\Big],  \label{18}
\end{eqnarray}
\end{widetext}
where $\mathcal{P}_t=\mathcal{P}_{\theta }=\mathcal{P}_{\phi }$. In the conventional approach, one may specify the energy density profile and subsequently determine the corresponding shape function $b(r)$. By imposing a particular equation of state (EoS), characterized by a chosen parameter $w$, the redshift function can then be obtained. In this work, rather than assuming a specific EoS, we begin with prescribed wormhole geometries characterized by different redshift functions and derive the corresponding equation of state that relates the pressure components to the energy density. This approach enables us to identify the effective EoS parameter associated with a given wormhole configuration. For notational simplicity, we set the reduced Planck constant equal to unity throughout this paper, namely $\hbar=1$.

Following Ref.~\cite{Garattini:2019ivd}, we promote the plates distance $a$ as a radial variable $r$. Using Eq.~(\ref{mo1}), the energy density can be recast in the compact form
\begin{equation}
\varrho_{I,II}=\varrho_{\rm \small{VED}}
\left[1+C^{(1)}_{I,II}\Bigg(\frac{\sqrt{\beta}}{r}\Bigg)^{2}+C^{(2)}_{I,II}\Bigg(\frac{\sqrt{\beta}}{r}\Bigg)^{4}\right]\,,\label{Mode1}
\end{equation}
where we have defined new parameters:
\begin{eqnarray}
C^{(1)}_{I}=\frac{10 \pi ^2}{9}\,,\,\,C^{(2)}_{I}=\frac{29 \pi^4}{15},\ C^{(1)}_{II}=\frac{5}{42} \left(3 \sqrt{10}+28\right) \pi ^2,\,\, C^{(2)}_{II}=\frac{1}{32} \left(93 \sqrt{10}+529\right) \pi ^4.\label{Cs}
\end{eqnarray}
%%%%%%%%%%%%%%%%%%%%
\subsection{Model with $\Phi(r)={\rm const.}$}\label{sec3A}
\begin{figure*}[t]
\centering
\includegraphics[height=.3\linewidth]{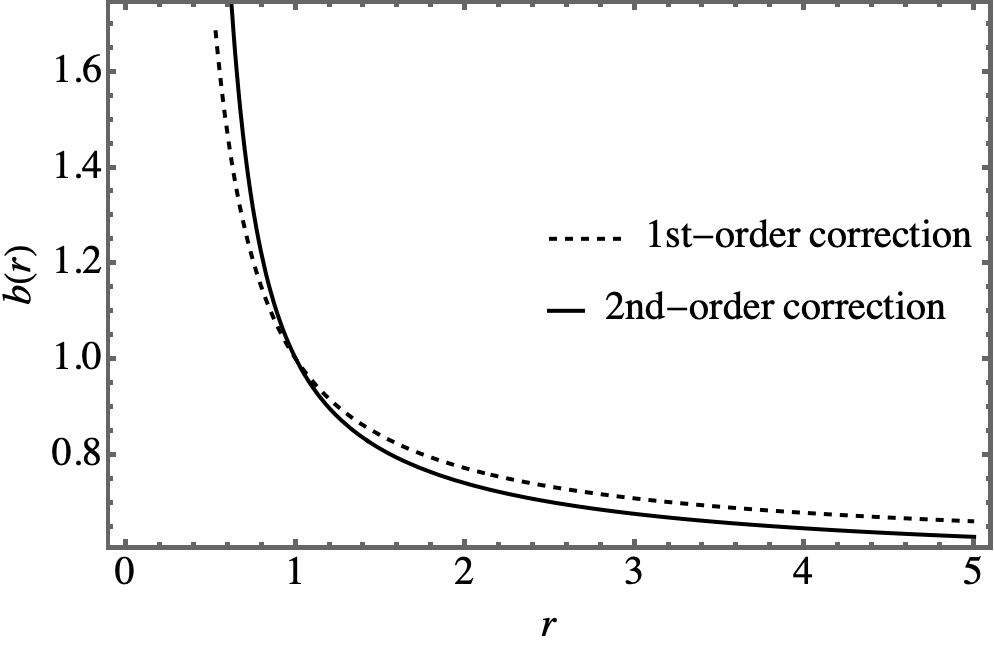}
\includegraphics[height=.3\linewidth]{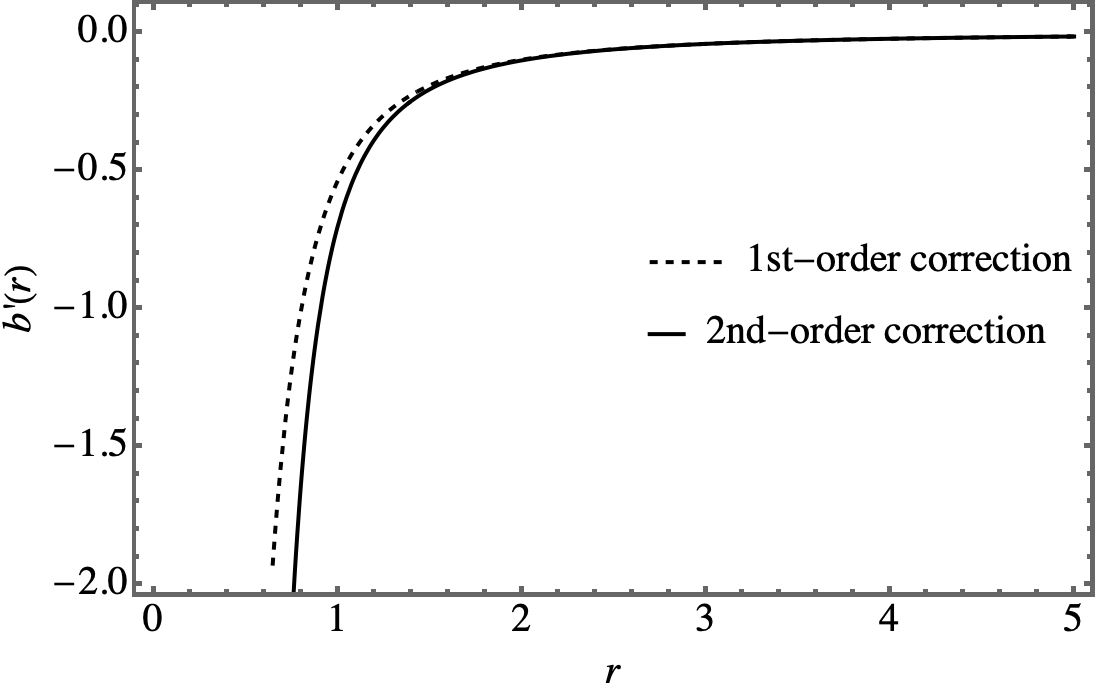}
\caption{We depict the shape function, $b_{I}(r)$, including corrections up to second order in the GUP parameter $\beta$, as a function of the radial coordinate $r$ (left panel). The corresponding derivative, $b'(r)$, is shown in the right panel to examine the fulfillment of the flare-out condition at the wormhole throat. Second-order GUP corrections preserve the fundamental
wormhole geometry while producing noticeable modifications
near the throat.
The derivative remains below unity,
confirming that the flare-out condition is satisfied. Throughout the analysis, we set $r_0=1$, $\hbar=1$, and $\beta=0.05$.}
\label{Fig1Mo1}
\end{figure*}

%%%%%%%%%%%%%%%%%%%%
As a first example, we consider the simplest wormhole configuration characterized by a constant redshift function, $\Phi(r)=\mathrm{const.}$, for which $\Phi'(r)=0$. This choice corresponds to a zero-tidal-force wormhole and naturally leads to an asymptotically flat spacetime geometry. Substituting the energy density Eq.~(\ref{Mode1}) into the field equation Eq.~(\ref{rho}) and integrating, we obtain the shape function
\begin{equation}
b_{I,II}(r)=C_1+\frac{\pi^3}{90}\Bigg[\frac{1}{r}+\frac{C^{(1)}_{I,II}\beta}{3 r^3}+\frac{C^{(2)}_{I,II}\beta^2}{5 r^5}\Bigg],
\end{equation}
where $C_1$ is an integration constant determined by the throat condition. We then use $b(r_0)=b_0=r_0$ to determine the constant $C_1$. Therefore, the shape function reads
\begin{eqnarray}
b_{I,II}(1)=r_0+\frac{\pi^3}{90}\Bigg[\Bigg(\frac{1}{r}-\frac{1}{r_{0}}\Bigg)+ \frac{C^{(1)}_{I,II}}{3}\Bigg(\frac{1}{r^3}-\frac{1}{r^{3}_{0}}\Bigg)\beta+\frac{C^{(2)}_{I,II}}{5}\Bigg(\frac{1}{r^5}-\frac{1}{r^{5}_{0}}\Bigg)\beta^2\Bigg]\,.\label{ShapI}
\end{eqnarray}
The second-order GUP correction contributes an additional
$r^{-5}$ term to the shape function.
Although this correction is subleading far from the throat,
it becomes increasingly important in the vicinity of
$r=r_0$, where quantum-gravity effects are expected to be strongest.
Consequently, the geometry receives localized corrections while
preserving asymptotic flatness. Clearly, in the limit $r\to \infty$, the asymptotically flat metric can be visualized in Fig.~\ref{Fig1Mo1}:
\begin{equation}
\lim_{r\to \infty}\frac{b_{I,II}(r)}{r}=0\,.
\end{equation}

%-------------------
\begin{figure}
\includegraphics[width=8 cm]{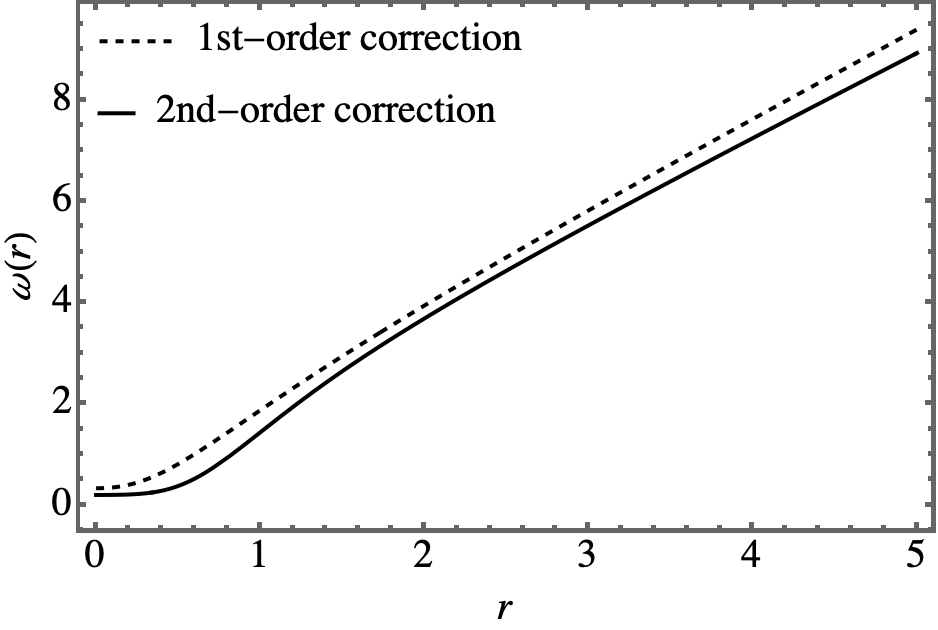}
\caption{We plot the equation-of-state (EoS) parameter $w_{I}(r)$ as a function of the radial coordinate $r$ for a GUP-corrected wormhole geometry characterized by a vanishing redshift function, $\Phi(r)=0$. The effective equation-of-state parameter increases
monotonically with radial distance.
The second-order correction slightly suppresses
$w(r)$ near the throat,
indicating stronger quantum-vacuum effects in the
high-curvature region. In our numerical analysis, we fix $r_0=1$, $\hbar=1$, and $\beta=0.05$.}\label{Fig1OmeM}
\end{figure}

\begin{figure*}[t]
\centering
\includegraphics[height=.3\linewidth]{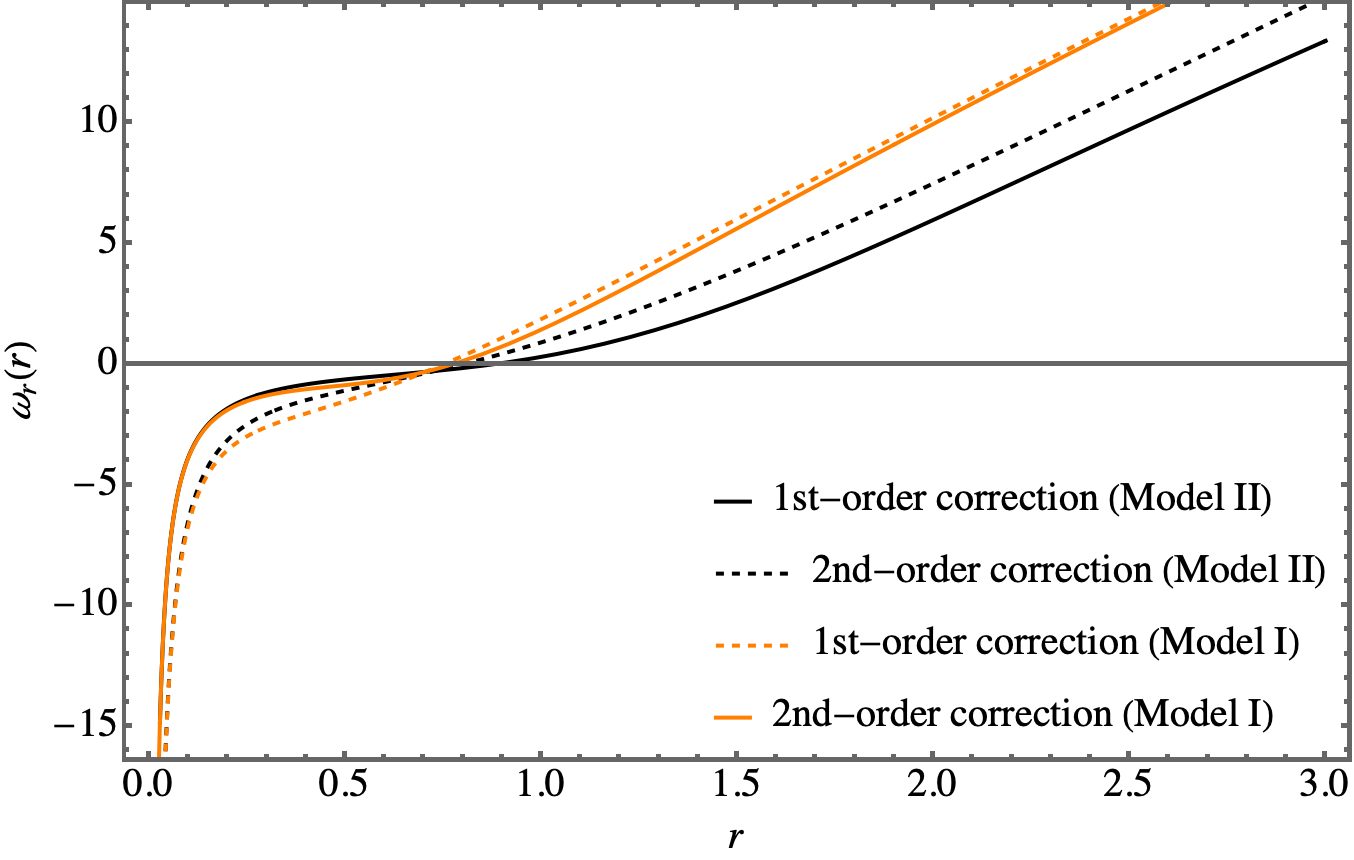}
\includegraphics[height=.3\linewidth]{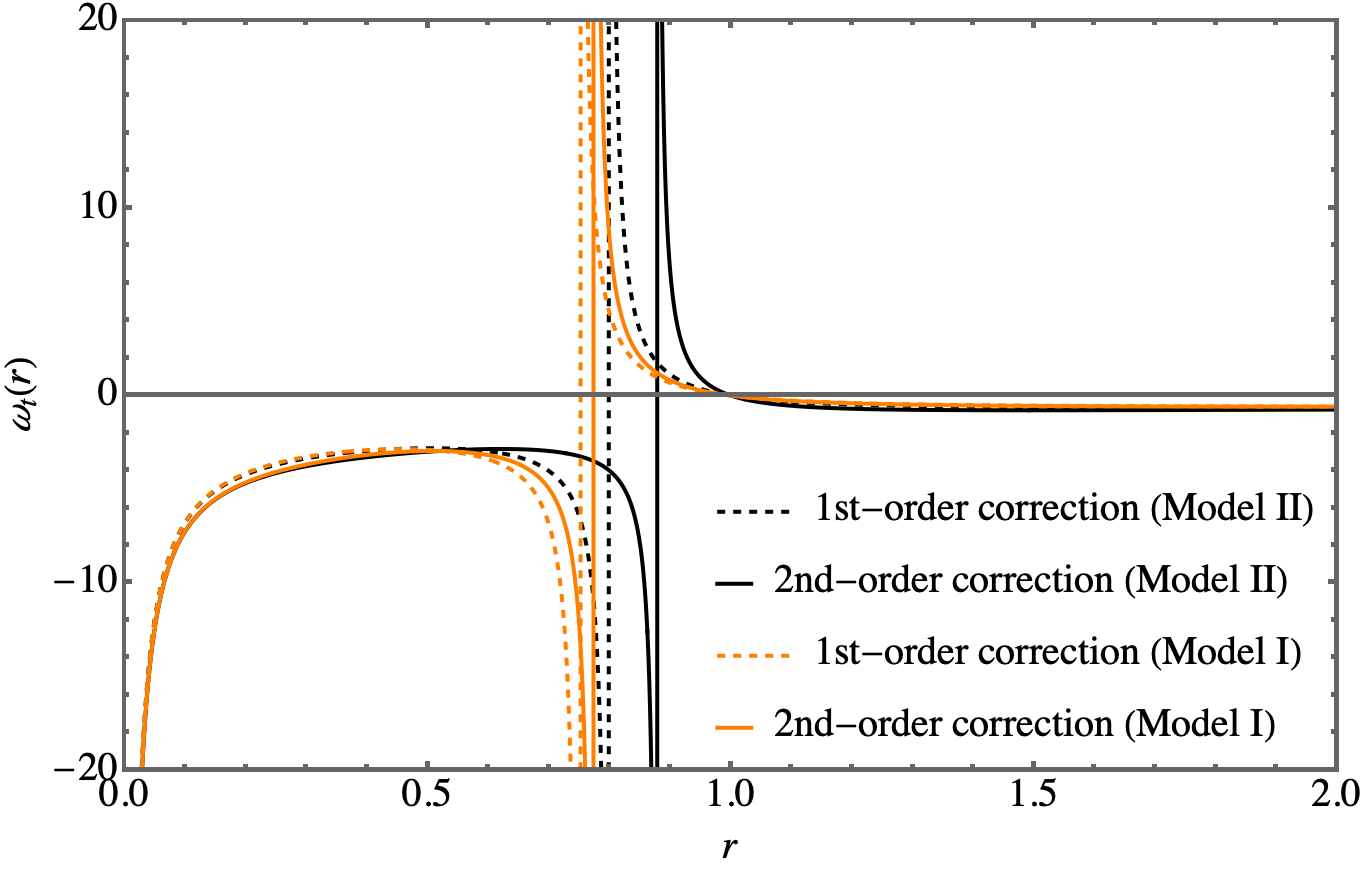}
\caption{We plot the equation-of-state (EoS) parameter $w^{I,II}_{r}(r)$ and $w^{I,II}_{t}(r)$ as a function of the radial coordinate $r$ for a GUP-corrected wormhole geometry characterized by a redshift function, $\Phi(r)=r_{0}/r$. The results include contributions up to second order in the GUP parameter. In our numerical analysis, we fix $r_0=1$, $\hbar=1$, and $\beta=0.05$.}\label{Fig1OmeMo1}
\end{figure*}

%-------------------
We plot in Fig.~\ref{Fig1OmeM} the equation-of-state (EoS) parameter $w(r)$ as a function of the radial coordinate $r$ for a GUP-corrected wormhole geometry characterized by a vanishing redshift function, $\Phi(r)=0$. The results include contributions up to second order in the GUP parameter. Employing the equation of state (EoS) relation,
\(
\mathcal{P}_r(r)=w(r)\rho(r),
\)
and considering the case of a vanishing redshift function, $\Phi(r)=0$, which corresponds to a tideless wormhole geometry, Eq.~(\ref{Pr}) reduces to
\begin{equation}\label{24}
8\pi r^{3}w_{I,II}(r)\rho_{I,II}(r)+b_{I,II}(r)=0.
\end{equation}
This expression can be solved straightforwardly for the EoS parameter, yielding
\begin{eqnarray}
w^{\Phi={\rm const.}}_{r,(I,II)}(r)\approx w^{(0)}_{r,(I,II)}(r)+w^{(1)}_{r,(I,II)}(r)\beta+w^{(2)}_{r,(I,II)}(r)\beta^{2}\,,
\end{eqnarray}
where a precise form of $w^{\Phi={\rm const.}}_{r,(I,II)}(r)$ is given in Appendix \ref{ApA}.
The left panel of Fig.~\ref{Fig1OmeMo1} illustrates the behavior of the shape function $b(r)$, incorporating contributions up to second order in the GUP parameter $\beta$, as a function of $r$. To verify the geometric requirements for a traversable wormhole, the right panel presents the derivative $b'(r)$, which is used to assess the flare-out condition.

%%%%%%%%%%%%%%%%%%%%
\subsection{Model with $\Phi(r)=r_{0}/r$}\label{sec3B}
%%%%%%%%%%%%%%%%%%%%
For the model with $\Phi(r)=r_{0}/r$, we find from Eq.~(\ref{Pr}) that
\begin{eqnarray}
\frac{b(r) r-2\,b(r)\,r_{0}+8 \pi  r^4 \rho(r) w_{r}(r) +2 r r_{0}}{8 \pi  r^4}=0\,.
\end{eqnarray}
Finally using the shape function (\ref{ShapI}) for the EoS parameter, we can compute $w(r)$ and write it in the form:
\begin{eqnarray}
w^{\Phi=r_{0}/r}_{r,(I,II)}(r)=w^{(0)}_{r,(I,II)}(r)+w^{(1)}_{r,(I,II)}(r)\beta+w^{(2)}_{r,(I,II)}(r)\beta^{2}\,,
\end{eqnarray}
where a precise form of $w^{\Phi=r_{0}/r}_{r,(I,II)}(r)$ is given in Appendix \ref{ApB}. Next, consider the scenario in which the equation of state takes the form
$P_t(r) = w_t(r)P_r(r)$, where $w_t(r)$ represents an arbitrary
function of $r$. Under these conditions, substituting the second equation
into the third in Eq.~(\ref{18}) yields the following expression
\begin{eqnarray}
&&r \Big\{2 r \big[r-b(r)\big] \Phi''(r)+\Phi'(r) \Big(2 r \big[r-b(r)\big] \Phi '(r)\nonumber\\&&+b(r) \big[4 w_{t}(r)-1\big]-4 r w_{t}(r)+2 r\Big)\Big\}+2 b(r) w_{t}(r)+ b(r)=0\,.
\end{eqnarray}
Therefore, the solution for $w^{I,II}_{t}(r)$ takes a form similar to that of $w^{\Phi=r_{0}/r}_{r,(I,II)}(r)$:
\begin{eqnarray}
w^{\Phi=r_{0}/r}_{t,(I,II)}(r)\approx w^{(0)}_{t,(I,II)}(r)+w^{(1)}_{t,(I,II)}(r)\beta+w^{(2)}_{t,(I,II)}(r)\beta^{2}\,,
\end{eqnarray}
where the precise form of $w^{\Phi=r_{0}/r}_{t,(I,II)}(r)$ is given in Appendix \ref{ApC}.

%%%%%%%%%%%%%%%%%%%%
\subsection{Isotropic model with $w_{r}(r)= {\rm const}.$}
%%%%%%%%%%%%%%%%%%%%
Starting from the conservation law of the energy--momentum tensor,
$\nabla_{\mu}T^{\mu\nu}=0$, one obtains the hydrostatic equilibrium equation governing the matter distribution that supports the wormhole geometry,
\begin{equation}
    \mathcal{P}'_r(r)=\frac{2\big[\mathcal{P}_t(r)-\mathcal{P}_r(r)\big]}{r}
    -\big[\rho(r)+\mathcal{P}_r(r)\big]\Phi'(r).
\end{equation}
Assuming an isotropic fluid configuration, $\mathcal{P}_t=\mathcal{P}_r$, together with the barotropic equation of state
\begin{equation}
    \mathcal{P}_r(r)=w_r\,\rho(r),
\end{equation}
where $w_r$ is taken to be a constant parameter, the above relation simplifies to
\begin{equation}
    w_r\,\rho'(r)
    =-(1+w_r)\rho(r)\,\Phi'(r),
\end{equation}
with the energy density $\rho(r)$ given by Eq.~(\ref{mo1}). Integrating the above equation, we find the corresponding redshift function
\begin{eqnarray}
    \Phi_{I,II}(r)=C+\frac{w_r}{1+w_r}
    \ln\!\left[
    \frac{r^{8}}
    {r^{4}+\beta C_{I,II}^{(1)}r^{2}+\beta^{2}C_{I,II}^{(2)}}
    \right],
    \label{iso}
\end{eqnarray}
where $C$ is an integration constant. The constant $C$ can be absorbed into the metric through a rescaling of the temporal coordinate, $dt\rightarrow C\,dt$. It is straightforward to verify that the redshift function remains finite at the throat, $r=r_0$, provided that $w_r\neq -1$. However, for large values of $r$, $\Phi(r)$ grows without bound, implying that the resulting spacetime is not asymptotically flat. Consequently, GUP-corrected wormhole solutions sustained by isotropic matter distributions cannot, in general, describe physically viable asymptotically flat wormholes.

%%%%%%%%%%%%%%%%%%%%
\subsection{Anisotropic model with $w_{r}(r)= {\rm const}.$}
%%%%%%%%%%%%%%%%%%%%
As discussed above, the isotropic configuration possesses only limited physical relevance. We therefore turn our attention to a more general anisotropic GUP-corrected Casimir wormhole solution. To this end, we adopt the equations of state
\begin{equation}
    \mathcal{P}_r(r)=w_r\,\rho(r), \qquad
    \mathcal{P}_t(r)=n\,w_r\,\rho(r),
\end{equation}
where $n$ is a constant parameter characterizing the degree of anisotropy. Substituting these relations into the conservation equation yields
\begin{equation}
    w_r\,\rho'(r)
    =
    \frac{2w_r(n-1)\rho(r)}{r}
    -(1+w_r)\rho(r)\Phi'(r).
\end{equation}
Integrating the above expression, we obtain the redshift function
\begin{eqnarray}
    \Phi_{I,II}(r)=C + \frac{w_r}{1+w_r}
    \ln\!\left[\frac{r^{2(n+3)}}{r^{4}+\beta C_{I,II}^{(1)}r^{2}+\beta^{2}C_{I,II}^{(2)}}\right],
\end{eqnarray}
where $C$ is an integration constant. It is worth noting that the isotropic solution given in Eq.~(\ref{iso}) is recovered in the special case $n=1$. As before, the solution becomes singular for $w_r=-1$. In contrast to the isotropic configuration, the anisotropic model admits asymptotically flat geometries. One can verify that $n=-1$ is the unique choice leading to asymptotic flatness.
\begin{figure}
\includegraphics[width=8 cm]{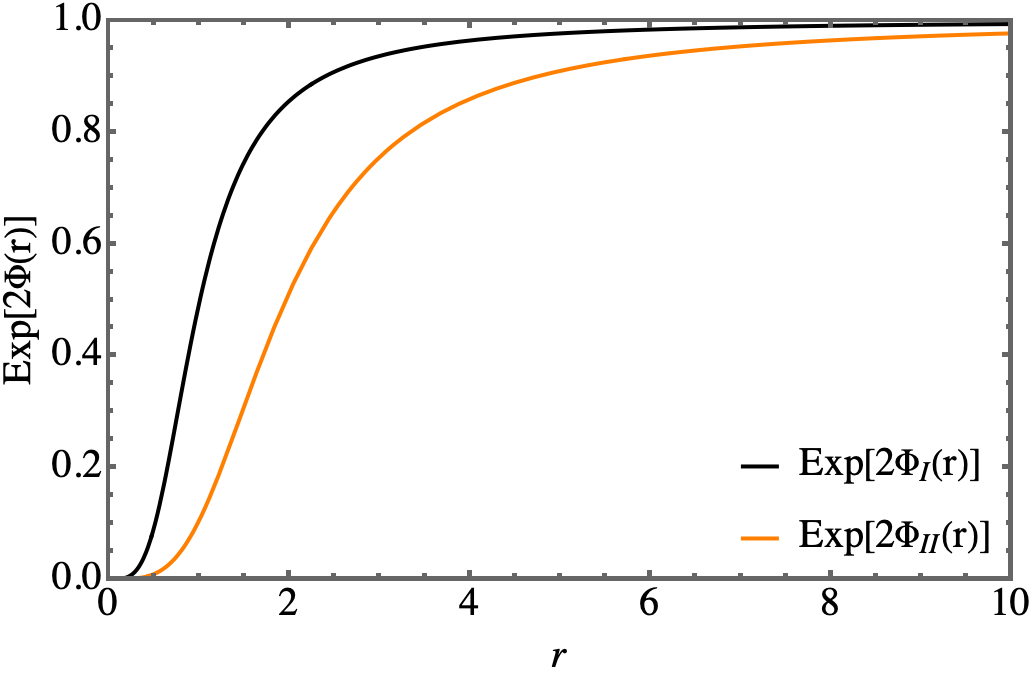}
\caption{We plot $\exp(2\Phi_{I,II}(r))$ as a function of the radial coordinate $r$ for the anisotropic case. We have used $r_0=1$, $\hbar=1$, $\beta=0.05$, $n=-1$ and $w=1$.}\label{Phi12}
\end{figure}

%%%%%%%%%%%%%%%%%%%%
\section{ADM mass for GUP-corrected Casimir
wormhole}\label{sec4}
%%%%%%%%%%%%%%%%%%%%
Now we evaluate the ADM mass of the GUP-corrected Casimir wormhole. To this end, we consider the asymptotically flat spatial geometry described by
\begin{eqnarray}
ds^2_{\Sigma} = \psi(r)dr^2+r^2\chi(r)\left(d\theta^2+\sin^2\theta d\phi^2\right),
\end{eqnarray}
where
\begin{equation}
\psi(r)=\frac{1}{1-\frac{b(r)}{r}}, \qquad \chi(r)=1.
\end{equation}
The ADM mass is evaluated using the standard expression (see, e.g., Ref.~\cite{Shaikh:2018kfv})
\begin{equation}\label{for}
m_{\rm ADM}=\lim_{r\to \infty}\frac{1}{2}\left[-r^2\chi'+r(\psi-\chi)\right].
\end{equation}
Substituting the above metric functions into Eq.~(\ref{for}) and taking the asymptotic limit yields
\begin{eqnarray}\label{ADM}
m_{I,II}^{\rm ADM}&=&\frac{r_0}{2}-\frac{\pi^3}{180 r_{0}}-\frac{\pi^3 C^{(1)}_{I,II}\beta}{540 r_{0}^3}-\underbrace{\frac{\pi^3 C^{(2)}_{I,II}\beta ^2}{900 r_{0}^5}}_{\rm a\,new\,contribution}.
\end{eqnarray}
This result shows that higher-order GUP effects systematically decrease the ADM mass. Physically, the second-order correction acts as an additional negative vacuum-energy contribution, reducing the effective
gravitational mass measured by distant observers. The four contributions in Eq.~(\ref{ADM}) have distinct physical origins. The firs-three terms have been reported in Ref.~\cite{Jusufi:2020rpw}, while the last-two terms represent the additional quantum-gravitational correction induced by the GUP. Regarding the GUP parameter, it is worth mentioning that Ref.~\cite{Das:2008kaa} discussed the possibility of placing phenomenological upper bounds on the quantum-gravity parameter by requiring consistency with experimental constraints at the electroweak scale.

%%%%%%%%%%%%%%%%%%%%
\section{Energy conditions}\label{sec5}
%%%%%%%%%%%%%%%%%%%%
Using the above results, we now examine the energy conditions and illustrate their behavior through the corresponding plots. The weak energy condition (WEC) requires that $T_{\mu\nu}u^{\mu}u^{\nu}\geq0$, where $T_{\mu\nu}$ denotes the energy-momentum tensor and $u^{\mu}$ is any timelike vector. For the present anisotropic matter distribution, this condition reduces to
\begin{equation}
\rho(r)+\mathcal{P}_{r}(r)\geq0.
\end{equation}
Likewise, the null energy condition (NEC) is expressed as
$T{\mu\nu}k^{\mu}k^{\nu}\geq0$, with $k^{\mu}$ representing a null vector, which similarly yields
\begin{equation}
\rho(r)+\mathcal{P}_{r}(r)\geq0.
\end{equation}
The strong energy condition (SEC) further requires
\begin{equation}
\rho(r)+2\mathcal{P}_{t}(r)\geq0,
\end{equation}
\begin{figure*}[t]
\centering
\includegraphics[height=.3\linewidth]{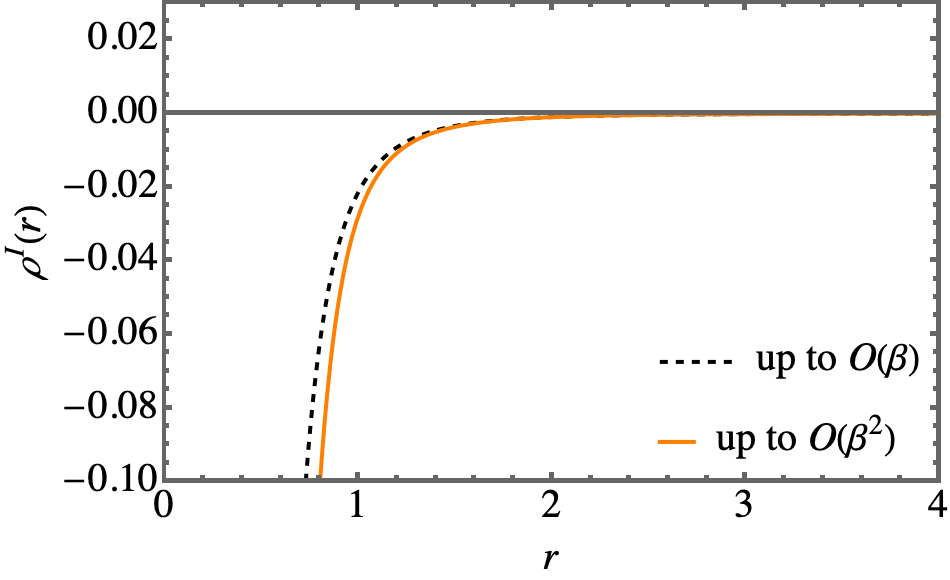}
\includegraphics[height=.3\linewidth]{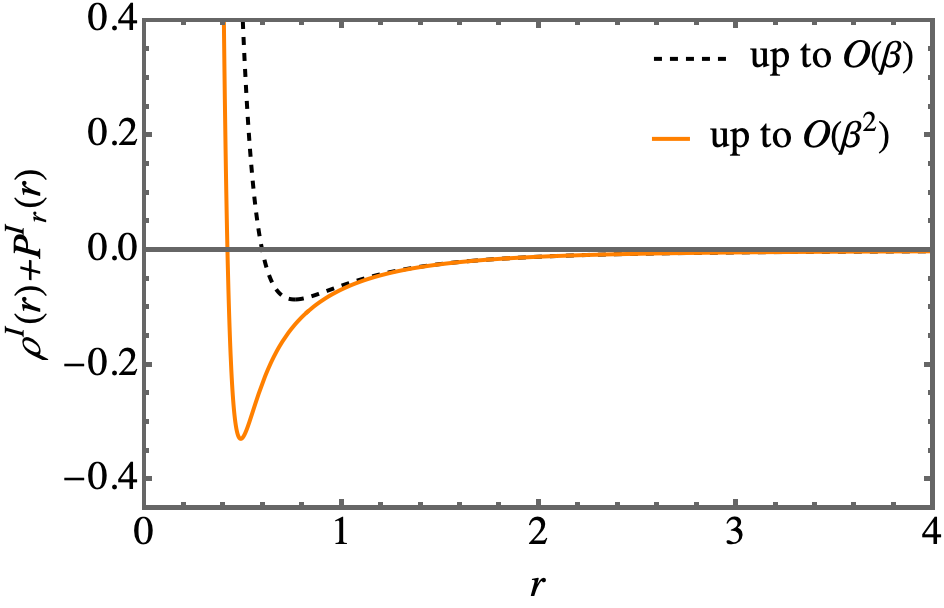}
\includegraphics[height=.3\linewidth]{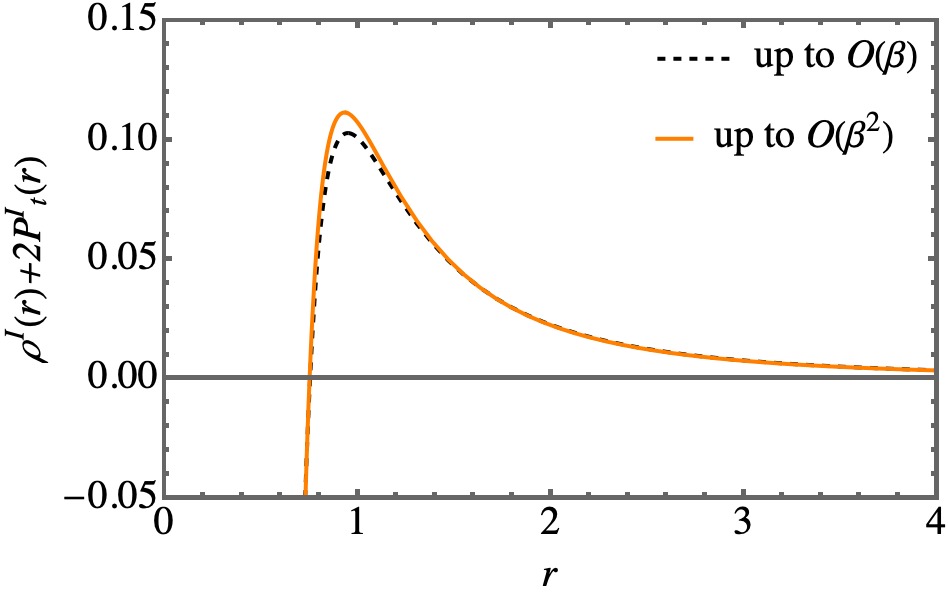}
\includegraphics[height=.3\linewidth]{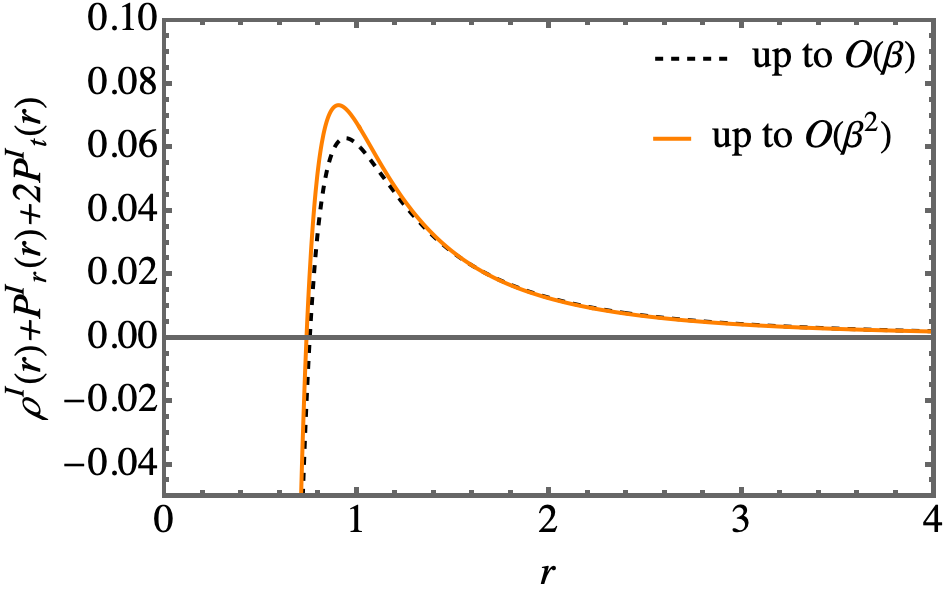}
\caption{The variation of the energy conditions incorporating the second-order GUP contributions as a function of $r$ using $\exp(2\Phi(r))=r_{0}/r$. Second-order GUP corrections produce only small
quantitative modifications to the classical energy
conditions.
The NEC and WEC remain violated near the throat,
while the magnitude of the violation remains localized. We use $r_0=1$, $\hbar=1$ and $\beta=0.05$.}\label{figEn}
\end{figure*}
together with
\begin{equation}
\rho(r)+\mathcal{P}_{r}(r)+2\mathcal{P}_{t}(r)\geq0.
\end{equation}
The behavior of these quantities is displayed in Fig.~\ref{figEn}. It is evident that the WEC and NEC are violated in the vicinity of the wormhole throat, $r=r_0$. More specifically, numerical evaluation shows that at $r_0=1$, $\left(\rho+\mathcal{P}_{r}\right)\big|_{r_{0}=1}<0$, although the violations remain quantitatively small.
From the perspective of quantum field theory, however, such violations are not unexpected. Quantum fluctuations are known to violate the classical energy conditions and may therefore provide the exotic matter necessary to support traversable wormholes. In this context, it is useful to consider the Quantum Weak Energy Condition (QWEC), which imposes the bound~\cite{Garattini:2019ivd}
\begin{equation}
\rho(r)+\mathcal{P}_{r}(r)<f(r), \qquad f(r)>0,
\end{equation}
where $r\in[r_0,\infty)$. Consequently, the small violations of the classical energy conditions observed in our solutions remain compatible with quantum field theory and may be interpreted as arising from quantum fluctuations rather than signaling any pathological behavior. The analysis demonstrates that introducing second-order
GUP corrections does not alter the qualitative behavior
of the energy conditions.
Instead, higher-order terms mainly affect the magnitude
of the violations while preserving the physical mechanism
supporting the traversable wormhole.

%%%%%%%%%%%%%%%%%%%%
\section{Amount of exotic matter}\label{sec6}
\begin{figure*}[t]
\centering
\includegraphics[height=.3\linewidth]{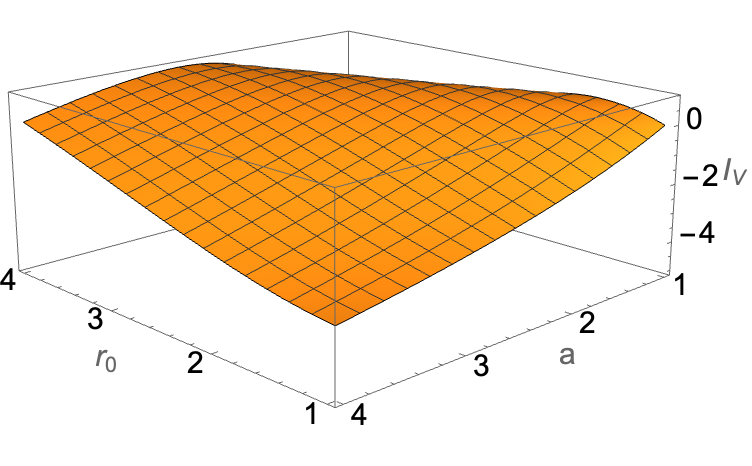}
\includegraphics[height=.3\linewidth]{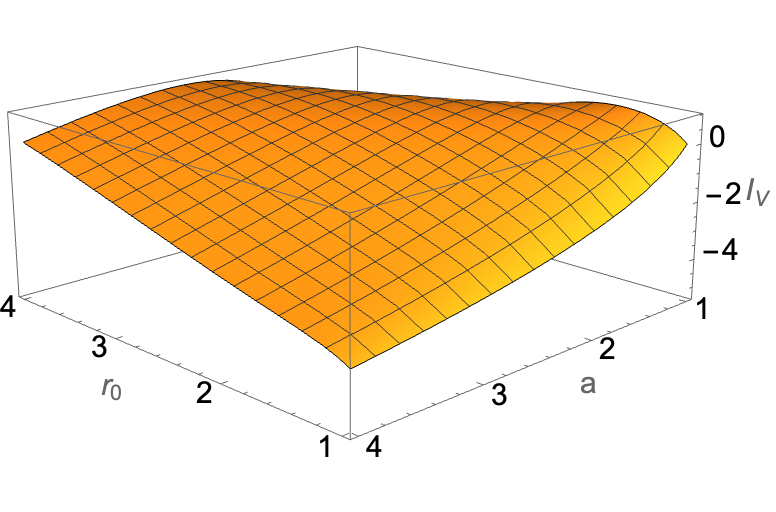}
\caption{The variation of $I_V$ against $r$ and $a$ of the case $\Phi=r_0/r$: Model I (left panel) and Model II (right panel). We use $r_0=1$, $\hbar=1$ and $\beta=0.05$.}\label{figIV}
\end{figure*}

%%%%%%%%%%%%%%%%%%%%
In this section, we briefly examine the volume integral quantifier, which provides a measure of the total amount of exotic matter required to sustain the wormhole geometry. This quantity depends exclusively on the energy density, $\rho$, and the radial pressure, $\mathcal{P}_r$, while remaining independent of the transverse pressure components. It is defined through the volume integral
\begin{eqnarray}
I_V=\oint [\rho+\mathcal{P}_r]~\mathrm{d}V
=2 \int_{r_{0}}^{\infty} \left( \rho+\mathcal{P}_r\right)~\mathrm{d}V,
\end{eqnarray}
which can equivalently be expressed as
\begin{eqnarray}
I_V =8 \pi \int_{r_{0}}^{\infty} \left( \rho+\mathcal{P}_r\right)~r^{2}\mathrm{d}r.
\end{eqnarray}
As discussed previously, the value of this integral characterizes the overall amount of exotic matter contained within the wormhole spacetime. We now evaluate this quantity for the shape function $b(r)$. To facilitate the calculation, we introduce a finite cutoff radius $a$, assuming that the wormhole interior extends from the throat radius $r_0$ to $r=a$. The volume integral then reduces to
\begin{equation}
I_V=8 \pi \int_{r_{0}}^{a} \left( \rho+\mathcal{P}_r\right)~r^{2}\mathrm{d}r.
\end{equation}
In the limiting case $a \rightarrow r_0$, the integral satisfies $I_{V} \rightarrow 0$, indicating that the total amount of exotic matter approaches zero as the integration region shrinks to the throat. In particular, for the redshift function $\Phi=r_0/r$, the Casimir wormhole can be sustained by an arbitrarily small quantity of exotic matter. Evaluating the above integral yields $I^V_{I,II}$:
\begin{eqnarray}
 I^V_{I,II}\approx I^{V,(0)}_{I,II}+I^{V,(1)}_{I,II}\beta+I^{V,(2)}_{I,II}\beta^{2}\,,
\end{eqnarray}
where a full expression for $I^V_{I,II}$ is given in Appendix \ref{ApD}. The logarithmic contribution reflects the cumulative effect
of the quantum vacuum over the wormhole interior, whereas
the polynomial terms originate from local GUP corrections.
The second-order contribution therefore modifies both the
local and global distributions of exotic matter. From Fig.~\ref{figIV}, adopting the redshift function $\exp(2\Phi(r))=r_{0}/r$, we calculate the volume integral quantifier $I_V$, which provides an estimate of the total exotic matter content. In both Models I and II, the analysis yields $I_V<0$, confirming that the corresponding traversable wormhole configurations require only a finite and comparatively small amount of exotic matter. Such a modest departure from the classical energy conditions is consistent with the possibility that the supporting negative energy originates from quantum fluctuations.

%%%%%%%%%%%%%%%%%%%%
\section{Theoretical Observational Signatures}
\label{sec:obs}

\subsection{Wormhole embedding}
\label{sec:embedding}

The embedding diagram gives a direct geometric picture of the wormhole throat. Restricting the metric of Eq.~\eqref{ShapI} to the equatorial slice $\theta=\pi/2$ and to a fixed time $t$ leaves a two-dimensional spatial section,
\begin{equation}
ds^{2}=\left(1-\frac{b(r)}{r}\right)^{-1}dr^{2}+r^{2}d\phi^{2},
\label{eq:embedmetric}
\end{equation}
with $b(r_{0})=r_{0}$ fixing the throat radius. This surface can be embedded in three-dimensional Euclidean space through
\begin{equation}
x=r\cos\phi,\qquad y=r\sin\phi,\qquad z=z(r),
\end{equation}
whose induced metric,
\begin{equation}
ds^{2}=\left[1+\left(\frac{dz}{dr}\right)^{2}\right]\,dr^{2}+r^{2}d\phi^{2},
\end{equation}
matches Eq.~\eqref{eq:embedmetric} once
\begin{equation}
\frac{dz}{dr}=\pm\left[\frac{r}{b(r)}-1\right]^{-1/2},\qquad z(r_{0}+\epsilon)=0,
\label{eq:embeddingDE}
\end{equation}
where $\epsilon$ is a small positive offset that keeps the numerical integration away from the coordinate singularity at $r=r_{0}$. Eq.~\eqref{eq:embeddingDE} diverges at the throat, so the embedded surface develops a vertical tangent there, the signature of the wormhole's minimal-radius bottleneck. Integrating outward,
\begin{equation}
z(r)=\pm\int_{r_{0}}^{r}\left[\frac{r'}{b(r')}-1\right]^{-1/2}dr',
\label{eq:zint}
\end{equation}
gives the full profile; flatness far from the throat requires $dz/dr\to0$ as $r\to\infty$, which holds automatically once $b(r)/r\to0$.

\begin{figure*}[t]
\centering
\includegraphics[height=.25\linewidth]{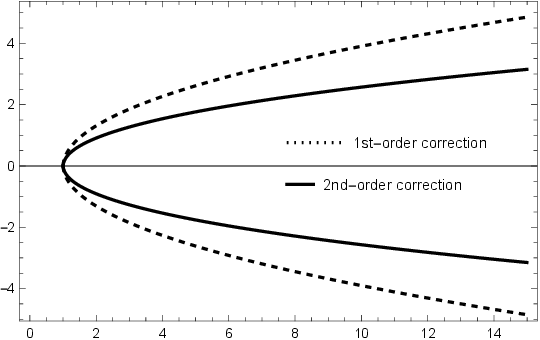}
\includegraphics[height=.27\linewidth]{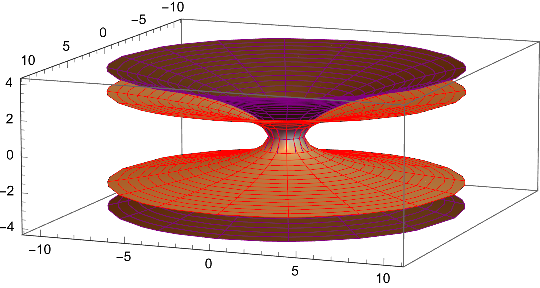}
\caption{Embedding diagram for $r_{0}=1$, $\beta=0.2$, as a function of the radial coordinate $r$, for the GUP-corrected wormhole with vanishing redshift function $\Phi(r)=0$. Curves include first- and second-order corrections in the GUP parameter.}
\label{Fig:7}
\end{figure*}

Fig.~\ref{Fig:7} shows the embedding surface obtained by numerically integrating Eq.~\eqref{eq:zint} with $r_{0}=1$ and $\beta=0.2$, for the shape function of Eq.~\eqref{ShapI} at $\Phi(r)=0$. Both sheets flare outward from the throat and flatten at large $r$, consistent with the asymptotic condition on $dz/dr$. The two curves in the left panel, corresponding to the first- and second-order GUP corrections, stay close together away from the throat but separate near $r=r_{0}$, where the $\beta^{2}$ term in $b(r)$ has the largest relative weight. Same in the right panel, upper circle shows the first-order, and lower circle shows the second order GUP correction.

\subsection{Gravitational lensing}
\label{sec:lensing}

\begin{figure*}[t]
\centering
\includegraphics[width=0.88\linewidth]{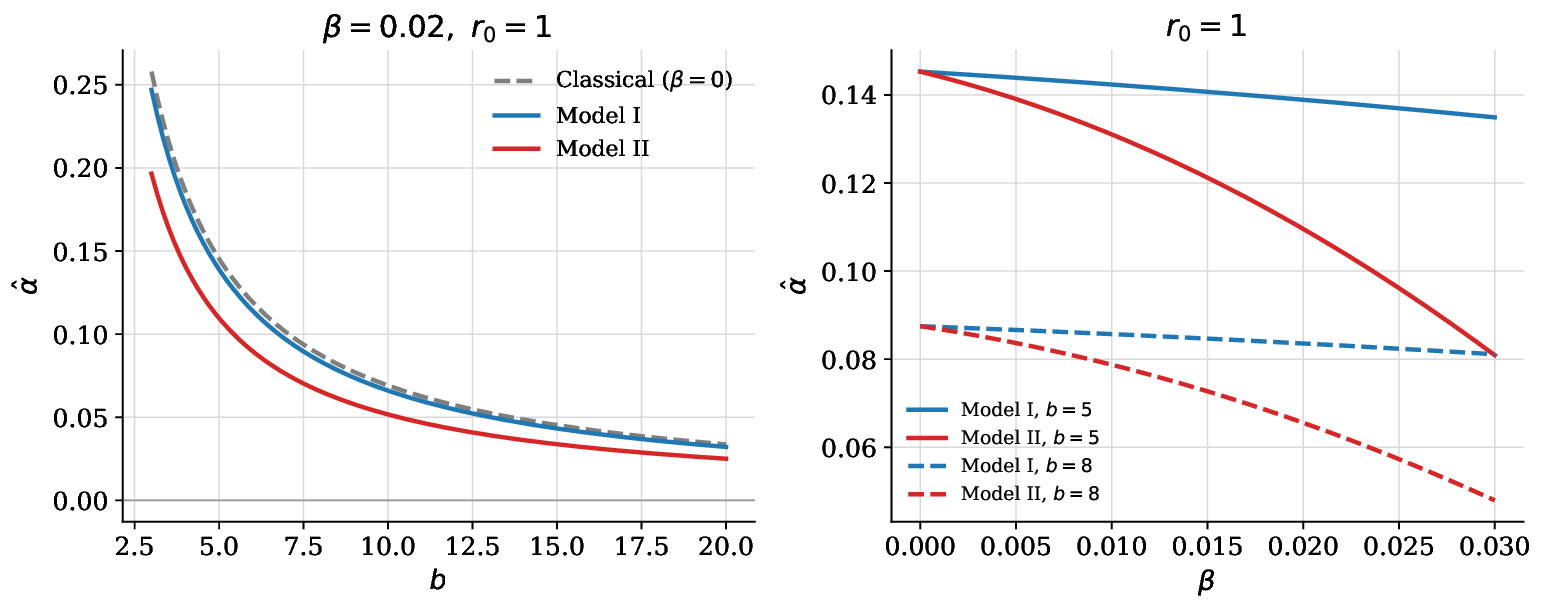}
\caption{Weak-field deflection angle $\hat\alpha$ from Eq.~\eqref{eq:defl_final}, for $r_{0}=1$. Left: $\hat\alpha$ against impact parameter $b$ at fixed $\beta=0.02$, comparing the classical ($\beta=0$) curve with Models I and II. Right: $\hat\alpha$ against $\beta\in[0,0.03]$ at fixed $b=5$ and $b=8$, for both models. The plotted $\beta$ range is chosen to stay within the region where the truncated series remains positive for both models; see text.}
\label{Fig:8}
\end{figure*}

We compute the weak-field deflection angle for the GUP-corrected wormhole using the Gauss--Bonnet method of Gibbons and Werner~\cite{Gibbons:2008rj}. We work throughout with the $\Phi(r)=0$ (zero-tidal-force) branch of Sec.~\ref{sec3A}, for which $g_{tt}=-1$ identically and $g_{rr}=1/A(r)$ with $A(r)\equiv 1-b(r)/r$. Null geodesics confined to the equatorial plane obey $-\dot t^{2}+\dot r^{2}/A(r)+r^{2}\dot\varphi^{2}=0$; setting $ds^{2}=0$ at fixed $t$ gives the optical metric for this branch,
\begin{equation}
dt^{2}=\frac{dr^{2}}{A(r)}+r^{2}\,d\varphi^{2},
\label{eq:optmetric}
\end{equation}
i.e. $g^{\rm opt}_{rr}=1/A(r)$ and $g^{\rm opt}_{\varphi\varphi}=r^{2}$. The corresponding area element on the optical 2-surface is $dS=\dfrac{r}{\sqrt{A(r)}}\,dr\,d\varphi$. The Gaussian curvature of the optical metric~\eqref{eq:optmetric}, for a line element of the form $dr^{2}/A(r)+r^{2}d\varphi^{2}$, is given exactly by
\begin{equation}
\mathcal{K}(r)=-\frac{A'(r)}{2r}=\frac{r\,b'(r)-b(r)}{2r^{3}}\,,
\label{eq:gausscurv}
\end{equation}
with no series truncation required; substituting the shape function of Eq.~\eqref{ShapI} gives $\mathcal{K}(r)$ exactly to $\mathcal{O}(\beta^{2})$, reported in Eq.~\eqref{eq:Kexpand} in Appendix~\ref{ApE}. The deflection angle follows from the Gauss-Bonnet theorem applied to a region $D_{R}$ bounded by the light ray and a circular arc $C_{R}$ of radius $R\to\infty$,
\begin{equation}
\iint_{D_{R}}\mathcal{K}\,dS+\oint_{\partial D_{R}}\kappa\,dt+\sum_{i}\theta_{i}=2\pi\chi(D_{R}),
\end{equation}
with $\chi(D_{R})=1$ for a simply connected domain. In the limit $R\to\infty$ the exterior angles sum to $\pi$ and the geodesic curvature of $C_{R}$ satisfies $\kappa(C_{R})\,dt\to d\varphi$, so the theorem reduces to
\begin{equation}
\hat\alpha=-\int_{0}^{\pi}\int_{b/\sin\varphi}^{\infty}\mathcal{K}(r)\,\frac{r}{\sqrt{A(r)}}\,dr\,d\varphi,
\label{eq:defl_int}
\end{equation}
where $b$ is the impact parameter, $r/\sqrt{A(r)}$ is the area-element weight fixed by Eq.~\eqref{eq:optmetric}, and the straight-line approximation $r(\varphi)=b/\sin\varphi$ is used for the zeroth-order photon trajectory. Substituting Eq.~\eqref{eq:Kexpand} into Eq.~\eqref{eq:defl_int} and expanding to second order in $\beta$ and to $\mathcal{O}(b^{-2})$ in the weak-field (large impact parameter) expansion gives the deflection angle reported in Eq.~\eqref{eq:defl_final} in Appendix~\ref{ApF}; direct numerical evaluation of the double integral in Eq.~\eqref{eq:defl_int}, without truncating in $1/b$, agrees with Eq.~\eqref{eq:defl_final} to within a few percent over the range $b=3$--$20$ plotted in Fig.~\ref{Fig:8}.

Fig.~\ref{Fig:8} plots Eq.~\eqref{eq:defl_final} for $r_{0}=1$. The left panel fixes $\beta=0.02$ and compares the classical ($\beta=0$) curve with both GUP models against impact parameter $b$: both models fall monotonically below the classical curve, with Model II suppressed more strongly than Model I at the same $\beta$, consistent with its larger coefficients $C^{(1)}_{II}$, $C^{(2)}_{II}$ of Eq.~\eqref{Cs}. The right panel fixes $b=5$ and $b=8$ and varies $\beta$ over $0$--$0.03$: $\hat\alpha$ decreases with $\beta$ for both models, with Model II falling off substantially faster. We restrict the plotted range to $\beta\lesssim0.03$ because, beyond this, the truncated Model II series in Eq.~\eqref{eq:defl_final} approaches zero and then crosses into unphysical negative values (already at $\beta\approx0.05$--$0.08$ for $b\gtrsim8$): this signals the breakdown of the perturbative expansion for Model II rather than a genuine sign change in the deflection, and mirrors the stronger GUP sensitivity of Model II already seen in the gravitational-wave echo spectrum of Fig.~\ref{Fig:echoII}.

\section{Gravitational Wave Echoes}
\label{sec:echoes}

We now examine the gravitational-wave echo signal associated with the GUP-corrected Casimir wormhole for the zero-tidal-force branch, $\Phi(r)=0$. A massless scalar perturbation on the background of Eq.~\eqref{5}, decomposed as $\Psi=\psi(r)\,Y_{\ell m}(\theta,\varphi)\,e^{-i\omega t}/r$, obeys the wave equation $\Box\Psi=0$. Introducing the tortoise coordinate
\begin{equation}
\frac{dr_{\star}}{dr}=\frac{1}{\sqrt{A(r)}}\,,\qquad A(r)=1-\frac{b(r)}{r}\,,
\label{eq:tortoise}
\end{equation}
brings the radial equation to the Schr\"odinger-like form
\begin{equation}
\frac{d^{2}\psi}{dr_{\star}^{2}}+\left[\omega^{2}-V_{\ell}(r)\right]\psi=0\,,
\label{eq:master}
\end{equation}
with the effective potential
\begin{equation}
V_{\ell}(r)=\frac{A'(r)}{2r}+\frac{\ell(\ell+1)}{r^{2}}
=\frac{b(r)-r\,b'(r)}{2r^{3}}+\frac{\ell(\ell+1)}{r^{2}}\,,
\label{eq:potential}
\end{equation}
where $\ell$ is the multipole number of the perturbation. Substituting the shape function of Eq.~\eqref{ShapI}, the value of the potential at the throat is
\begin{equation}
V_{\ell}(r_{0})=\frac{\ell(\ell+1)}{r_{0}^{2}}+\frac{1}{2r_{0}^{2}}+\frac{\pi^{3}}{180\,r_{0}^{4}}+\frac{C^{(1)}_{I,II}\beta\,\pi^{3}}{180\,r_{0}^{6}}+\frac{C^{(2)}_{I,II}\beta^{2}\pi^{3}}{180\,r_{0}^{8}}\,,
\label{eq:Vthroat}
\end{equation}
and, for every $\ell=0,1,2,3$ and both GUP models over $\beta=0$--$0.5$, we find that $V_{\ell}(r)$ decreases monotonically from $V_{\ell}(r_{0})$ at the throat down through zero, i.e.\ $V_{\ell}'(r)<0$ on the entire interval where $V_{\ell}(r)\geq0$. (Beyond the point where $V_{\ell}$ first turns negative it develops a shallow minimum before returning to zero as $r\to\infty$; this feature lies entirely below zero and is irrelevant for real, positive-energy modes.) Consequently $V_{\ell}(r)$ has no positive local maximum away from the throat, unlike the geometric photon-sphere barrier of a spacetime with $\Phi'(r)\neq0$.

Because $V_{\ell}(r)$ has no off-throat maximum, the reflection of an incident wave is not the geometric-optics effect associated with a photon sphere; it is a purely wave-mechanical (tunnelling) effect governed by how far the potential exceeds $\omega^{2}$ outside the throat. For a mode of real frequency $\omega$ with $\omega^{2}<V_{\ell}(r_{0})$, the region $r_{0}<r<r_{b}$ is classically forbidden, where the turning point $r_{b}=r_{b}(\omega,\ell)$ is defined implicitly by
\begin{equation}
V_{\ell}(r_{b})=\omega^{2}\,.
\label{eq:turningpoint}
\end{equation}
Since $V_{\ell}(r)$ decreases monotonically and continuously from $V_{\ell}(r_{0})$ to $0$, Eq.~\eqref{eq:turningpoint} has a unique root $r_{b}(\omega,\ell)>r_{0}$ for every $\omega$ in the trapped band $0<\omega<\omega_{\max}\equiv\sqrt{V_{\ell}(r_{0})}$; for $\omega\geq\omega_{\max}$ the mode propagates freely at all $r\geq r_{0}$ and no reflection occurs. The wave is thus partially reflected at the steep potential wall near the throat and partially transmitted through the evanescent region $(r_{0},r_{b})$, so that a fraction of the incident amplitude is trapped and reflects repeatedly between the throat and $r_{b}(\omega,\ell)$ before leaking away. The associated round-trip echo delay follows from the tortoise coordinate of Eq.~\eqref{eq:tortoise},
\begin{equation}
t_{\rm echo}(\omega,\ell)\approx 2\int_{r_{0}}^{r_{b}(\omega,\ell)}\frac{dr}{\sqrt{A(r)}}\,,
\label{eq:techo}
\end{equation}
with $r_{b}(\omega,\ell)$ obtained from Eq.~\eqref{eq:turningpoint} together with Eq.~\eqref{eq:potential} and the shape function of Eq.~\eqref{ShapI}; because $V_{\ell}(r)$ is a rational function of $r$, Eq.~\eqref{eq:turningpoint} is transcendental in $r_{b}$ once GUP corrections are included, and $r_{b}(\omega,\ell)$ is obtained numerically for given $r_{0}$, $\beta$, and $\ell$.

Eq.~\eqref{eq:techo} differs qualitatively from the echo-delay formula of a spacetime with a genuine photon sphere. There, $r_{b}=r_{\rm ph}$ is fixed by the geometry alone, so all frequencies below the barrier top share a common, $\omega$-independent echo period. Here, by contrast, $r_{b}$ depends on $\omega$ through Eq.~\eqref{eq:turningpoint}: modes closer to $\omega_{\max}$ probe shallower turning points nearer the throat and produce shorter, more tightly spaced echoes, while modes well below $\omega_{\max}$ tunnel through more of the potential and produce longer delays, with $r_{b}\to r_{0}$ and $t_{\rm echo}\to0$ as $\omega\to\omega_{\max}$. The predicted echo signal for this class of GUP-corrected Casimir wormholes is therefore intrinsically dispersive, with a frequency-dependent delay set by Eqs.~\eqref{eq:potential}--\eqref{eq:techo} rather than by a single geometric time scale, and it vanishes altogether for $\omega\geq\omega_{\max}=\sqrt{V_{\ell}(r_{0})}$ of Eq.~\eqref{eq:Vthroat}.

\begin{figure*}[t]
\centering
\includegraphics[width=0.88\linewidth]{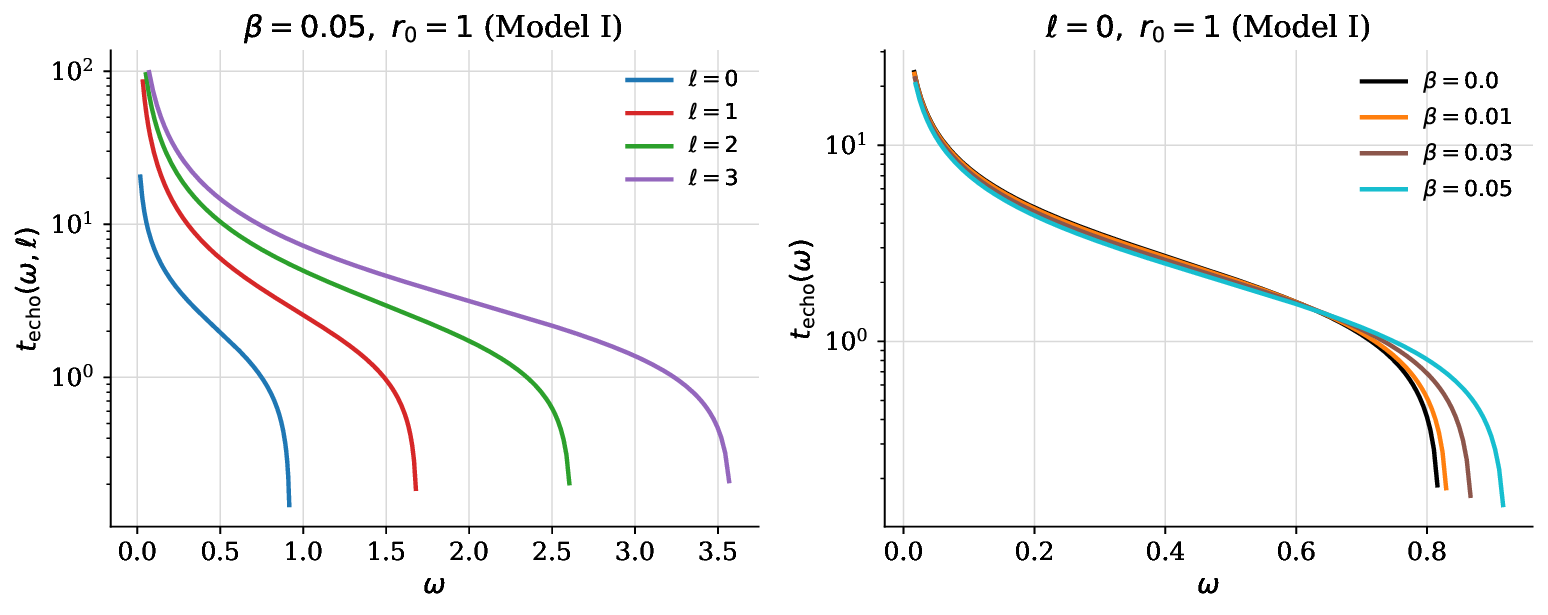}
\caption{Echo delay $t_{\rm echo}(\omega,\ell)$ from Eqs.~\eqref{eq:turningpoint} and \eqref{eq:techo}, for $r_{0}=1$ (Model I). Left: fixed $\beta=0.05$, multipoles $\ell=0,1,2,3$. Right: fixed $\ell=0$, GUP parameter $\beta=0,0.01,0.03,0.05$. In both panels $t_{\rm echo}\to0$ as $\omega\to\omega_{\max}=\sqrt{V_{\ell}(r_{0})}$ and diverges as $\omega\to0^{+}$.}
\label{Fig:echo}
\end{figure*}

\begin{figure*}[t]
\centering
\includegraphics[width=0.88\linewidth]{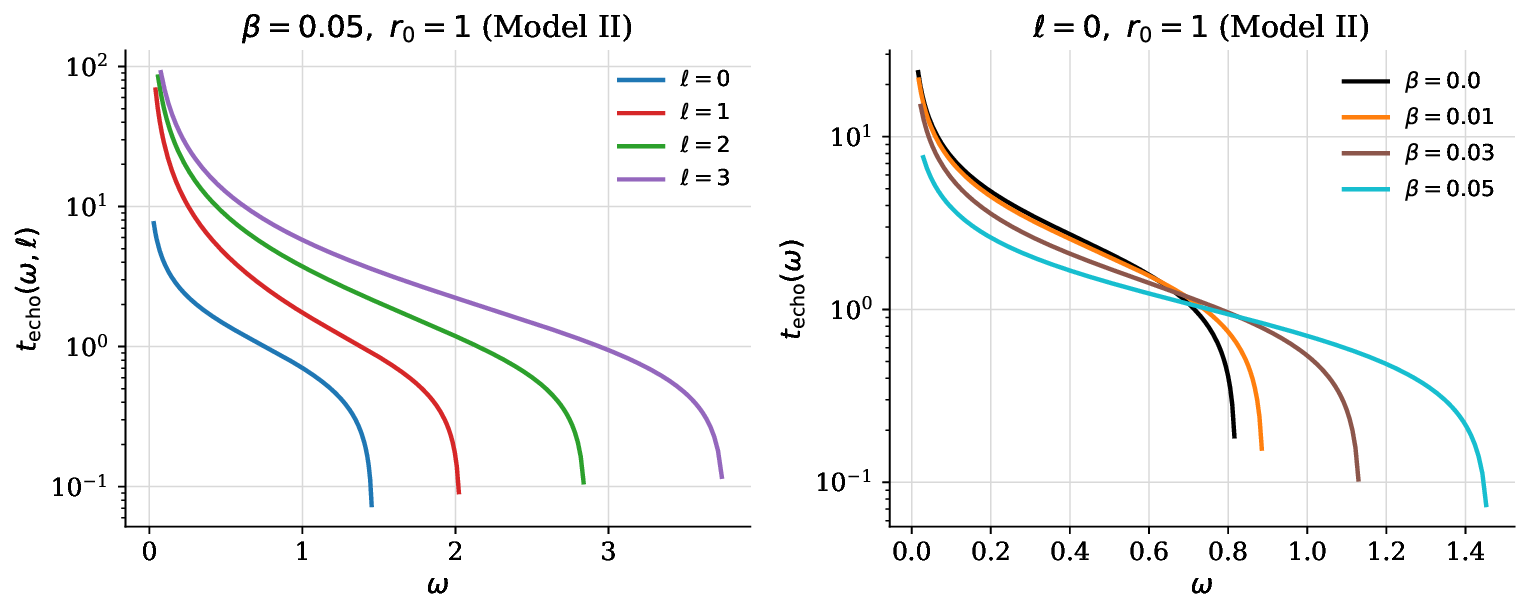}
\caption{Echo delay $t_{\rm echo}(\omega,\ell)$ from Eqs.~\eqref{eq:turningpoint} and \eqref{eq:techo}, for $r_{0}=1$ (Model II). Left: fixed $\beta=0.05$, multipoles $\ell=0,1,2,3$. Right: fixed $\ell=0$, GUP parameter $\beta=0,0.01,0.03,0.05$. As in Fig.~\ref{Fig:echo}, $t_{\rm echo}\to0$ as $\omega\to\omega_{\max}=\sqrt{V_{\ell}(r_{0})}$ and diverges as $\omega\to0^{+}$; the larger GUP coefficients of Model II shift $\omega_{\max}$ upward more sharply at the same $\beta$.}
\label{Fig:echoII}
\end{figure*}

Fig.~\ref{Fig:echo} plots $t_{\rm echo}(\omega,\ell)$ obtained by numerically solving Eq.~\eqref{eq:turningpoint} for $r_{b}(\omega,\ell)$ and evaluating Eq.~\eqref{eq:techo}, for $r_{0}=1$, for both GUP models. The left panel of each figure fixes $\beta=0.05$ and varies the multipole $\ell=0,1,2,3$: increasing $\ell$ raises the throat value $V_{\ell}(r_{0})$ of Eq.~\eqref{eq:Vthroat} and hence $\omega_{\max}$, shifting each curve to higher frequency, while all four curves show the same qualitative divergence as $\omega\to0$ and decay to zero as $\omega\to\omega_{\max}$. The right panel of each figure fixes $\ell=0$ and varies the GUP parameter over the smaller, perturbatively well-controlled range $\beta=0,0.01,0.03,0.05$: even this modest increase in $\beta$ raises $\omega_{\max}$ and shortens the delay at fixed $\omega$, showing that the GUP correction compresses the dispersive echo spectrum toward higher frequencies already at small $\beta$. Both panels use a logarithmic $t_{\rm echo}$ axis because the delay diverges as $\omega\to0^{+}$ (where $r_{b}\to\infty$). Comparing Fig.~\ref{Fig:echo} (Model I) with Fig.~\ref{Fig:echoII} (Model II) at the same $\beta$ values, Model II's larger coefficients $C^{(1)}_{II}$, $C^{(2)}_{II}$ of Eq.~\eqref{Cs} produce a visibly stronger upward shift of $\omega_{\max}$ and a correspondingly sharper compression of the echo spectrum for the same nominal $\beta$, consistent with the stronger GUP sensitivity of Model II already seen in the lensing and shadow-type observables of earlier sections.

%%%%%%%%%%%%%%%%%%%%%%%%%%%%%%%%%%%%%%%%%%%%
\section{Connecting the GUP parameter to phenomenological bounds}\label{sec:GUPbounds}
%%%%%%%%%%%%%%%%%%%%%%%%%%%%%%%%
In this section, we relate the GUP parameter employed in our
wormhole solutions to the dimensionless parameter commonly used in
phenomenological studies. This comparison requires particular care
because the parameter $\beta$ appearing in the modified commutation
relation is dimensional. In the present framework, the GUP algebra is
written schematically as
\begin{equation}
 [\hat{x},\hat{p}]
 =
 i\hbar\left(1+\beta \hat{p}^{\,2}+\cdots\right).
 \label{eq:GUPcommutator_bound}
\end{equation}
Since the combination $\beta \hat{p}^{\,2}$ must be dimensionless,
$\beta$ has dimensions of inverse momentum squared. The corresponding
minimum-length scale is
\begin{equation}
 \ell_{\rm GUP}
 \equiv
 \Delta x_{\min}
 =
 \hbar\sqrt{\beta}.
 \label{eq:GUP_length}
\end{equation}
It follows that $\hbar^{2}\beta$ has dimensions of length squared. Phenomenological analyses commonly introduce a dimensionless GUP parameter $\beta_{0}$ according to Ref.~\cite{Das:2008kaa}
\begin{equation}
 \beta
 =
 \frac{\beta_{0}}{(M_{\rm Pl}c)^{2}}
 =
 \frac{\beta_{0}\ell_{\rm Pl}^{2}}{\hbar^{2}},
 \label{eq:beta_beta0}
\end{equation}
where
\begin{equation}
 \ell_{\rm Pl}
 =
 \sqrt{\frac{\hbar G}{c^{3}}}
 \simeq
 1.616\times10^{-35}\ {\rm m}
 \label{eq:Planck_length}
\end{equation}
is the Planck length. Eqs.~(\ref{eq:GUP_length}) and
(\ref{eq:beta_beta0}) give
\begin{equation}
 \ell_{\rm GUP}
 =
 \sqrt{\beta_{0}}\,\ell_{\rm Pl}.
 \label{eq:GUP_length_beta0}
\end{equation}

For a wormhole geometry, the natural characteristic length is the
throat radius $r_{0}$. We therefore introduce the dimensionless
combination
\begin{equation}
 \bar{\beta}
 \equiv
 \frac{\hbar^{2}\beta}{r_{0}^{2}}
 =
 \left(\frac{\ell_{\rm GUP}}{r_{0}}\right)^{2}.
 \label{eq:betabar_definition}
\end{equation}
This quantity arises directly in the GUP-corrected Casimir energy.
Indeed, the correction terms can be written as
\begin{equation}
 \left(\frac{\hbar\sqrt{\beta}}{r}\right)^{2}
 =
 \frac{\bar{\beta}}{x^{2}},
 \qquad
 \left(\frac{\hbar\sqrt{\beta}}{r}\right)^{4}
 =
 \frac{\bar{\beta}^{2}}{x^{4}},
 \qquad
 x\equiv\frac{r}{r_{0}}.
 \label{eq:dimensionless_GUP_terms}
\end{equation}
Thus, after measuring all radial distances in units of $r_{0}$, the
actual expansion parameter is $\bar{\beta}$. Since our numerical
analysis adopts $r_{0}=1$ and $\hbar=1$, the numerical parameter
labelled $\beta$ in the figures coincides with $\bar{\beta}$. In other
words, a plotted value such as $\beta=0.05$ should be interpreted as
\begin{equation}
 \bar{\beta}
 =
 \frac{\hbar^{2}\beta}{r_{0}^{2}}
 =
 0.05,
 \label{eq:beta005}
\end{equation}
rather than as $\beta_{0}=0.05$. This value corresponds to
\begin{equation}
 \frac{\ell_{\rm GUP}}{r_{0}}
 =
 \sqrt{0.05}
 \simeq 0.224,
 \label{eq:length_ratio005}
\end{equation}
so that the minimum length is approximately $22.4\%$ of the throat
radius. Combining Eqs.~(\ref{eq:beta_beta0}) and
(\ref{eq:betabar_definition}) yields the conversion
\begin{equation}
 \bar{\beta}
 =
 \beta_{0}
 \left(\frac{\ell_{\rm Pl}}{r_{0}}\right)^{2}
 \label{eq:betabar_beta0}
\end{equation}
or, equivalently,
\begin{equation}
 \beta_{0}
 =
 \bar{\beta}
 \left(\frac{r_{0}}{\ell_{\rm Pl}}\right)^{2}.
 \label{eq:beta0_betabar}
\end{equation}
This relation shows explicitly that a numerical value of
$\bar{\beta}$ cannot be converted into a unique value of $\beta_{0}$
unless the physical throat radius is specified. The upper bounds have been found as Ref.~\cite{Das:2008kaa}:
\begin{align}
\beta_0 &< 10^{36} &&\text{(Lamb shift)}, \label{eq:lamb}\\
\beta_0 &< 10^{50} &&\text{(Landau levels)}, \label{eq:landau}\\
\beta_0 &< 10^{21} &&\text{(scanning tunnelling current)}, \label{eq:stm}
\end{align}
together with the theoretical requirement $\beta_0 \lesssim 10^{34}$ for consistency
with the non-observation of GUP effects at the electroweak scale. The case $\beta_0\sim1$
corresponds to genuinely Planckian, and hence experimentally inaccessible, corrections. This allows $\beta_0$ to be read off for any assumed throat size and compared directly
against the phenomenological bounds of Eqs.~\eqref{eq:lamb}, \eqref{eq:landau} and \eqref{eq:stm}, without
first solving for $r_0^{\max}$. Table~\ref{tab:beta0-direct} evaluates
Eq.~\eqref{eq:beta0_betabar} for $\tilde\beta = 0.05$ across a representative range of
throat sizes, from the Planck length itself up to macroscopic scales, and indicates
which of the four bounds each value is compatible with.

\begin{table}[t]
\centering
\caption{Predicted $\beta_0$ from Eq.~\eqref{eq:beta0_betabar} for $\tilde\beta = 0.05$,
as a function of assumed throat radius $r_0$, compared against the phenomenological
bounds of Ref.~\cite{Das:2008kaa}.}
\label{tab:beta0-direct}
\begin{tabular}{lccl}
\hline\hline
$r_0$ & $r_0/\ell_{Pl}$ & $\beta_0(r_0)$ & Compatible with \\
\hline
$\ell_{Pl}\;(1.6\times10^{-35}\,{\rm m})$ & $1$              & $0.05$        & all bounds \\
$1\,{\rm fm}\;(10^{-15}\,{\rm m})$        & $6.2\times10^{19}$ & $1.95\times10^{38}$ & Landau only \\
$1\,\text{\AA}\;(10^{-10}\,{\rm m})$      & $6.2\times10^{24}$ & $1.95\times10^{48}$ & Landau only \\
$1\,{\rm nm}\;(10^{-9}\,{\rm m})$         & $6.2\times10^{25}$ & $1.95\times10^{50}$ & none \\
$1\,{\rm mm}$                              & $6.2\times10^{31}$ & $1.95\times10^{62}$ & none \\
$1\,{\rm km}$                              & $6.2\times10^{37}$ & $1.95\times10^{74}$ & none \\
\hline\hline
\end{tabular}
\end{table}

The result is already at the femtometre scale, $\beta_0 \sim 10^{38}$ exceeds
the STM bound ($10^{21}$), the electroweak-consistency bound ($10^{34}$), and the Lamb
shift bound ($10^{36}$), remaining compatible only with the far weaker Landau-level
bound ($10^{50}$); by the \AA ngstrom scale even that bound is saturated, and beyond
$\sim 1\,$nm every phenomenological constraint in Ref.~\cite{Das:2008kaa} is violated for
$\tilde\beta = 0.05$. Only a throat within a few orders of magnitude of $\ell_{Pl}$
itself yields a $\beta_0$ safely below all four bounds simultaneously, reproducing the
$\beta_0 = \mathcal{O}(1)$ estimate of Eq.~\eqref{eq:beta0_betabar}.

\subsection{Independent experimental and astrophysical bounds}
\label{sec:betaboundstable}

The four bounds of Eqs.~\eqref{eq:lamb}, \eqref{eq:landau} and \eqref{eq:stm} are not the only constraints placed on the GUP deformation parameter. Table~\ref{tab:beta0bounds} collects representative bounds from laboratory mechanical-oscillator experiments, equivalence-principle tests, gravitational-wave dispersion timing, and black-hole shadow imaging, drawing on the compilation of Ref.~\cite{Scardigli2019}. These sources generally quote a bound on a dimensionless parameter denoted simply $\beta$ in the original literature; we identify this with $\beta_{0}$ of Eq.~\eqref{eq:beta_beta0} at the order-of-magnitude level used throughout this comparison, since an $\mathcal{O}(1)$ rescaling between different GUP realizations is possible and has not been worked out for Model~II specifically.

\begin{table}[t]
\centering
\caption{Representative upper bounds on the dimensionless GUP parameter $\beta_{0}$ (identified, at the order-of-magnitude level, with the correspondingly normalized parameter reported in each source) from independent experiments and observations.}
\label{tab:beta0bounds}
\begin{tabular}{lll}
\hline\hline
System / probe & Bound & Reference \\
\hline
Electroweak length-scale consistency & $\beta_{0}\lesssim10^{34}$ & \cite{Das:2008kaa} \\
Hydrogen Lamb shift & $\beta_{0}<10^{36}$ & \cite{Das:2008kaa} \\
Landau levels & $\beta_{0}<10^{50}$ & \cite{Das:2008kaa} \\
STM tunneling current & $\beta_{0}<10^{21}$ & \cite{Das:2008kaa} \\
Cryogenic micro/nano-oscillators & $\beta_{0}\lesssim10^{8\text{--}12}$ & \cite{Bawaj2015,Scardigli2019} \\
Sapphire mechanical resonator & $\beta_{0}\lesssim5.2\times10^{6}$ & \cite{Bushev2019} \\
Equivalence-principle violation (allowed) & $\beta_{0}\lesssim10^{21}$ & \cite{Ghosh2014} \\
GW150914 graviton-dispersion timing & $\beta_{0}<2.3\times10^{60}$ & \cite{Feng2017GW} \\
M87* shadow circularity (EHT) & $\beta_{0}<10^{90}$ & \cite{Neves2020} \\
\hline\hline
\end{tabular}
\end{table}

As with the four bounds of Eqs.~\eqref{eq:lamb}, \eqref{eq:landau} and \eqref{eq:stm}, the laboratory bounds in Table~\ref{tab:beta0bounds} sit many orders of magnitude above the theoretically favored $\beta_{0}\sim1$, while the gravitational-wave and shadow bounds are weaker still, since both probe $\beta_{0}$ through a small relativistic correction (graviton dispersion, shadow circularity) rather than a direct quantum-mechanical energy shift. None of the entries in Table~\ref{tab:beta0bounds} excludes $\beta_{0}=\mathcal{O}(1)$.

\subsection{Crossover throat radius}
\label{sec:crossover}

Eq.~\eqref{eq:beta0_betabar} shows that a bound $\beta_{0}<\beta_{0}^{\max}$ restricts $\bar\beta=\beta_0(\ell_{\rm Pl}/r_0)^2$ rather than $\beta_0$ alone: for a fixed bound, larger throats are pushed to smaller $\bar\beta$. Setting $\bar\beta=1$ in Eq.~\eqref{eq:betabar_beta0} defines a crossover throat radius
\begin{equation}
r_{0}^{*}=\ell_{\rm Pl}\sqrt{\beta_{0}^{\max}}\,,
\label{eq:crossover}
\end{equation}
below which a given bound still permits order-unity second-order GUP-Casimir corrections to the wormhole geometry of the kind computed in Secs.~\ref{sec3}--\ref{sec:echoes}, and above which it forces $\bar\beta\ll1$, making those corrections unobservably small. Table~\ref{tab:crossover} evaluates Eq.~\eqref{eq:crossover} for each bound of Eqs.~\eqref{eq:lamb}, \eqref{eq:landau}, \eqref{eq:stm} and Table~\ref{tab:beta0bounds}, using $\ell_{\rm Pl}=1.616\times10^{-35}\,{\rm m}$.

Two conclusions follow. First, for any wormhole throat of laboratory or astrophysical size ($r_{0}\gg10^{-10}\,{\rm m}$), every bound in Eqs.~\eqref{eq:lamb}, \eqref{eq:landau}, \eqref{eq:stm}, and Table~\ref{tab:beta0bounds} except the shadow bound already forces the second-order Casimir corrections of Eqs.~\eqref{mo1} and~\eqref{fmod2} to be far below any conceivable detection threshold, consistent with the generic expectation that Planck-suppressed effects are unmeasurable away from the Planck scale itself. Second, the current shadow-based bound is weak enough that it does not by itself rule out $\mathcal{O}(1)$ GUP-Casimir corrections even for a throat as large as $r_{0}^{*}\sim2\times10^{10}\,{\rm m}$ -- of order twenty times the solar radius -- leaving this regime open only because no dedicated shadow, lensing, or echo analysis of a horizonless Casimir-supported compact object has yet been performed.

\begin{table}[t]
\centering
\caption{Crossover throat radius $r_{0}^{*}$, Eq.~\eqref{eq:crossover}, below which each bound on $\beta_{0}$ permits order-unity second-order GUP-Casimir corrections to the wormhole geometry.}
\label{tab:crossover}
\begin{tabular}{lc}
\hline\hline
Bound source & $r_{0}^{*}$ \\
\hline
Sapphire resonator ($\beta_{0}\lesssim5.2\times10^{6}$) & $\sim4\times10^{-32}\,{\rm m}$ \\
Cryogenic oscillators ($\beta_{0}\lesssim10^{12}$) & $\sim2\times10^{-29}\,{\rm m}$ \\
STM tunneling current ($\beta_{0}<10^{21}$) & $\sim5\times10^{-25}\,{\rm m}$ \\
Electroweak consistency ($\beta_{0}\lesssim10^{34}$) & $\sim2\times10^{-18}\,{\rm m}$ \\
Lamb shift ($\beta_{0}<10^{36}$) & $\sim2\times10^{-17}\,{\rm m}$ \\
Landau levels ($\beta_{0}<10^{50}$) & $\sim2\times10^{-10}\,{\rm m}$ \\
GW150914 dispersion ($\beta_{0}<2.3\times10^{60}$) & $\sim2\times10^{-5}\,{\rm m}$ \\
M87* shadow (EHT, $\beta_{0}<10^{90}$) & $\sim2\times10^{10}\,{\rm m}$ \\
\hline\hline
\end{tabular}
\end{table}
The observables derived in Secs.~\ref{sec:lensing} and~\ref{sec:echoes} for the present ($\Phi=0$) branch are, in fact, sensitive to exactly this regime. Figure~\ref{Fig:8} shows the weak-field deflection angle $\hat\alpha$ falling measurably below its classical value once $\beta\gtrsim0.01$--$0.02$, with Model II responding several times more strongly than Model I at the same $\beta$; Figs.~\ref{Fig:echo} and~\ref{Fig:echoII} show the dispersive echo delay $t_{\rm echo}(\omega)$ and cutoff frequency $\omega_{\max}$ shifting in the same direction. Either observable could, in principle, be used the way Ref.~\cite{Neves2020} used M87*'s shadow circularity to bound $\beta_{0}$ for a Schwarzschild black hole: a measured deflection angle or echo spectrum for a horizonless compact object would translate, via Eq.~\eqref{eq:beta0_betabar}, into a bound on $\beta_{0}$ at that object's physical throat scale. We are not aware of a dedicated search of this kind for a Casimir-wormhole candidate, and we leave it, together with a corresponding analysis for the $\Phi'(r)\neq0$ branch where a genuine photon sphere exists, to future work.

%%%%%%%%%%%%%%%%%%%%%%%%%%%%%%%%
\section{Conclusions}\label{sec9}
%%%%%%%%%%%%%%%%%%%%%%%%%%%%%%%%
In this work, we have investigated traversable wormholes supported by Casimir vacuum energy incorporating second-order corrections arising from the generalized uncertainty principle (GUP). Extending two representative GUP models beyond the previously known first-order approximation, we derived analytical expressions for the Casimir energy density and pressure up to $\mathcal{O}(\beta^{2})$ and employed them as the matter source in the Einstein field equations. The resulting solutions demonstrate that second-order quantum-gravity corrections systematically modify the wormhole geometry while preserving the essential requirements for traversability, including the existence of a throat, the flare-out condition, and asymptotic flatness for appropriate choices of the redshift function.

The higher-order GUP corrections were shown to produce additional contributions to the shape function, effective equation of state, and ADM mass. In particular, the second-order terms reduce the effective gravitational mass of the wormhole and alter the amount of exotic matter required to support the throat. An analysis of the energy conditions reveals that the null and weak energy conditions remain violated only within a small neighborhood of the throat, while the corresponding volume integral quantifier confirms that the total amount of exotic matter remains finite and can be made arbitrarily small. These features are consistent with the interpretation that the required exotic matter originates from quantum vacuum fluctuations rather than classical matter sources.

An important outcome of the present analysis is that the second-order corrections reveal the emergence of a systematic perturbative structure for the GUP-modified Casimir vacuum energy. Rather than representing isolated corrections, the first- and second-order contributions suggest that the Casimir energy can be organized as the expansion
\begin{equation}
{\cal E}=\mathcal{E}_{\rm \small{VED}}\left[
1+\sum_{k} C_{k}
\left(
\frac{\hbar\sqrt{\beta}}{a}
\right)^{2k}
\right],
\end{equation}
where $\mathcal{E}_{\rm \small{VED}}$ denotes the standard Casimir energy, while the coefficients $C_{k}$ depend on the particular realization of the generalized uncertainty principle. The results presented here determine the first two nontrivial coefficients for two distinct GUP models, providing evidence that quantum-gravitational corrections possess an underlying hierarchical structure. This perturbative framework offers a natural starting point for systematically constructing higher-order corrections and exploring their convergence properties.

We have also investigated theoretical observational signatures of the resulting wormhole geometries. The embedding diagrams confirm the geometric consistency of the solutions, the weak-field gravitational lensing calculation of Sec.~\ref{sec:lensing} exhibits systematic dependence on the GUP parameter, and the gravitational-wave echo analysis of Sec.~\ref{sec:echoes} shows that, in the absence of a photon sphere, trapping and reflection proceed through a purely wave-mechanical mechanism that yields a frequency-dependent echo delay, vanishing above a model-dependent cutoff frequency $\omega_{\max}$.

To relate the parameter range employed in the wormhole analysis to existing phenomenological constraints, we define the dimensionless GUP parameter as $\bar{\beta}=\beta/r_0^2$, where $r_0$ denotes the throat radius. Using the conventional parametrization $\beta=\beta_0\ell_{\rm Pl}^2$, the dimensionless phenomenological parameter becomes $\beta_0=\bar{\beta}(r_0/\ell_{\rm Pl})^2$. Therefore, an upper bound $\beta_0\leq\beta_0^{\rm max}$ requires $\bar{\beta}\leq\beta_0^{\rm max}(\ell_{\rm Pl}/r_0)^2$, or equivalently, $r_0\leq\ell_{\rm Pl}\sqrt{\beta_0^{\rm max}}$ once $\bar\beta$ is set to order unity, defining the crossover throat radius $r_0^{*}$ of Sec.~\ref{sec:crossover}. This relation shows that the numerical values of $\bar{\beta}$ cannot be compared directly with phenomenological bounds unless the throat radius is specified. Collecting bounds on $\beta_0$ from laboratory quantum systems, mechanical-oscillator and equivalence-principle experiments, gravitational-wave dispersion timing, and black-hole shadow imaging (Sec.~\ref{sec:betaboundstable}), we find that every bound except the current EHT shadow constraint already forces $r_0^{*}$ well below any laboratory or astrophysical scale, so that order-unity second-order GUP-Casimir corrections remain observationally viable only for near-Planckian throats. The one exception -- the shadow bound, with $r_0^{*}\sim2\times10^{10}\,{\rm m}$ -- leaves room for a macroscopic Casimir wormhole with $\mathcal{O}(1)$ corrections, and the weak-field deflection angle (Sec.~\ref{sec:lensing}) and dispersive echo spectrum (Sec.~\ref{sec:echoes}) derived here are, in principle, the observables through which such a bound could be set directly for a horizonless compact object, in the same spirit as existing shadow- and dispersion-based bounds on black holes. Hence, the values used in the plots represent the relative strength of the GUP corrections, while their physical interpretation depends explicitly on the throat scale $r_0$.

In summary, the present work demonstrates that second-order GUP corrections not only provide quantitative refinements to Casimir-supported wormholes but also establish a systematic framework for investigating minimal-length effects in self-gravitating quantum vacuum configurations. The analytical formalism developed here provides a foundation for studying higher-order quantum-gravity corrections and may ultimately contribute to bridging fundamental quantum-gravity models with future astrophysical observations.

Several directions deserve further investigation. One immediate extension is to derive higher-order coefficients $C_k$ in the perturbative expansion of the GUP-corrected Casimir energy and investigate whether universal relations emerge among different GUP models or whether the resulting series remains intrinsically model dependent. It would also be worthwhile to construct rotating, e.g., Refs.~\cite{Teo:1998dp,Garattini:2025gfq,Nedkova:2013msa,Gyulchev:2018fmd,Tangphati:2023uxt,Errehymy:2026urb}, and electrically charged, e.g., Refs.~\cite{Samart:2021tvl,Koga:2025bqw,Garattini:2025cix}, GUP-corrected Casimir wormholes, where frame dragging and electromagnetic fields are expected to produce richer lensing phenomenology, or even consider the consequences of finite temperature contributions to a traversable wormhole, see e.g., Refs.~ \cite{Garattini:2024jkr,Muniz:2025teu}. A photon sphere and shadow, unlike the wave-mechanical echo signal of Sec.~\ref{sec:echoes}, require a redshift function with $\Phi'(r)\neq0$ outside the throat -- e.g. the $\Phi(r)=r_0/r$ branch of Sec.~\ref{sec3B} or another non-constant choice -- and revisiting these observables with such a $\Phi(r)$, using the corrected optical-metric formalism of Sec.~\ref{sec:lensing}, is left for future work. Another promising direction is to extend the present framework to modified theories of gravity or to dynamical wormhole spacetimes, allowing one to explore the interplay between quantum-gravity corrections and cosmological evolution. On the observational side, a detailed analysis of quasinormal modes, gravitational-wave ringdown signals, and echoes, together with comparisons with observations from the Event Horizon Telescope and future gravitational-wave detectors such as LISA and the Einstein Telescope, may provide phenomenological constraints on the GUP parameter. Such investigations would further clarify whether higher-order quantum-gravity corrections can leave observable imprints on compact objects and traversable wormholes.

%%%%%%%%%%%%%%%%%%%%%%%%%%%%%%%%
\section*{Acknowledgments}\label{sA}
The work of Kazuharu Bamba was supported in part by the JSPS KAKENHI Grants No. 24KF0100 and No. 25KF0176, as well as by a Grant-in-Aid for Academic Research from the Yamaguchi Scholarship Foundation. Phongpichit Channuie acknowledges financial support from the National Science, Research and Innovation Fund (NSRF) through the Program Management Unit for Human Resources \& Institutional Development, Research and Innovation (PMU-B) under Grant No. B39G690007.

\appendix

%%%%%%%%%%%%%
\section{EoS for $\Phi={\rm cont.}: w^{I,II}_{r}(r)$}
\label{ApA}
%%%%%%%%%%%%%%%%%%
In this appendix, we present the explicit analytical expression for the radial equation-of-state parameter and briefly clarify its physical origin. The radial equation-of-state (EoS) parameter corresponding to a constant redshift
function, $\Phi(r)=\mathrm{const.}$, can be obtained directly from the Einstein
field equations by combining the energy density with the radial pressure.
For this choice of the redshift function, the spacetime is free from tidal
forces, and therefore the matter properties are encoded entirely in the
shape function and the corresponding EoS parameter.
The resulting expression for $w_r(r)\equiv P_r/\rho$ is
\begin{eqnarray}
w^{\Phi={\rm const.}}_{r,(I,II)}(r)
&=&\frac{1}{15 \pi ^3 r_{0}^5 \left(\beta C^{(1)}_{I,II}r^2+\beta ^2 C^{(2)}_{I,II}+r^4\right)}\Bigg[\pi^3 \Big(5 \beta  C^{(1)}_{I,II} r^2 r_{0}^2 \left(r_{0}^3-r^3\right)+3 \beta^2 C^{(2)}_{I,II} \left(r_{0}^5-r^5\right)\notag\\&&\qquad\qquad\qquad\qquad\qquad\qquad\qquad\qquad+15 r^4 r_{0}^4 (r_{0}-r)\Big)+1350 r^5 r_{0}^6\Bigg],
\label{e27}
\end{eqnarray}
where $C^{(1)}_{I,II}$ and $C^{(2)}_{I,II}$ denote the first- and second-order
GUP coefficients for Models I and II, respectively. The parameter $\beta$
characterizes the quantum-gravity correction associated with the generalized
uncertainty principle, while $r_0$ represents the wormhole throat radius.

Equation~(\ref{e27}) shows that the radial EoS is no longer a constant but
depends explicitly on the radial coordinate through polynomial functions of
$r$. Consequently, the effective matter supporting the wormhole is intrinsically
inhomogeneous, with its local thermodynamic properties varying throughout the
spacetime. The numerator contains contributions arising from the classical
Casimir background together with first- and second-order GUP corrections,
whereas the denominator is governed by the modified energy density. The
competition between these terms determines the radial evolution of
$w_r(r)$.

In the limit $\beta\rightarrow0$, all quantum-gravity corrections vanish,
reducing Eq.~(\ref{e27}) smoothly to the classical Casimir-supported
wormhole EoS. For finite $\beta$, however, both the magnitude and radial
dependence of $w_r(r)$ are modified, indicating that the GUP corrections
alter the effective pressure required to sustain the wormhole geometry.
These modifications become increasingly significant in the vicinity of the
throat, where the minimal-length effects are strongest, while at large
distances the corrections rapidly diminish and the classical behavior is
recovered.

%%%%%%%%%%%%%
\section{EoS for $\Phi=r_{0}/r: w^{I,II}_{r}(r)$}
\label{ApB}
%%%%%%%%%%%%%%%%%%
In this appendix, we describe the explicit form and radial dependence of the equation-of-state parameter associated with a non-constant redshift function. The radial equation-of-state (EoS) parameter corresponding to the non-constant
redshift function
\begin{equation}
\Phi(r)=\frac{r_0}{r},
\end{equation}
is obtained by combining the modified energy density and radial pressure
derived from the Einstein field equations. Unlike the constant-redshift case,
this choice introduces finite tidal forces while remaining regular throughout
the spacetime since $\Phi(r)\rightarrow0$ as $r\rightarrow\infty$. The corresponding EoS parameter, defined by
\begin{eqnarray}
w^{I,II}_{r}(r)&=&\frac{3 r^4}{\beta  C^{(1)}_{I,II} r^2+\beta ^2 C^{(2)}_{I,II}+r^4}+\frac{\beta  C^{(1)}_{I,II} r^2}{3 \left(\beta  C^{(1)}_{I,II} r^2+\beta ^2 C^{(2)}_{I,II}+r^4\right)}+\frac{\beta ^2 C^{(2)}_{I,II}}{5 \left(\beta  C^{(1)}_{I,II} r^2+\beta ^2 C^{(2)}_{I,II}+r^4\right)}\notag\\&&+\frac{2 \beta ^2 C^{(2)}_{I,II} r^4}{5 r_{0}^4 \left(\beta  C^{(1)}_{I,II} r^2+\beta ^2 C^{(2)}_{I,II}+r^4\right)}-\frac{180 r^4 r_{0}^2}{\pi ^3 \left(\beta  C^{(1)}_{I,II} r^2+\beta ^2 C^{(2)}_{I,II}+r^4\right)}+\frac{2 \beta  C^{(1)}_{I,II} r^4}{3 r_{0}^2 \left(\beta  C^{(1)}_{I,II} r^2+\beta ^2 C^{(2)}_{I,II}+r^4\right)}\notag\\&&-\frac{2 \beta  C^{(1)}_{I,II} r r_{0}}{3 \left(\beta  C^{(1)}_{I,II} r^2+\beta ^2 C^{(2)}_{I,II}+r^4\right)}-\frac{2 \beta ^2 C^{(2)}_{I,II} r_{0}}{5 r \left(\beta  C^{(1)}_{I,II} r^2+\beta ^2 C^{(2)}_{I,II}+r^4\right)}-\frac{\beta ^2 C^{(2)}_{I,II} r^5}{5 r_{0}^5 \left(\beta  C^{(1)}_{I,II} r^2+\beta ^2 C^{(2)}_{I,II}+r^4\right)}\notag\\&&-\frac{\beta  C^{(1)}_{I,II} r^5}{3 r_{0}^3 \left(\beta  C^{(1)}_{I,II} r^2+\beta ^2 C^{(2)}_{I,II}+r^4\right)}+\frac{270 r^5 r_{0}}{\pi ^3 \left(\beta  C^{(1)}_{I,II} r^2+\beta ^2 C^{(2)}_{I,II}+r^4\right)}-\frac{r^5}{r_{0} \left(\beta  C^{(1)}_{I,II} r^2+\beta ^2 C^{(2)}_{I,II}+r^4\right)}\notag\\&&-\frac{2 r^3 r_{0}}{\beta  C^{(1)}_{I,II} r^2+\beta ^2 C^{(2)}_{I,II}+r^4}\,.\label{B2}
\end{eqnarray}
Eq.~(\ref{B2}) demonstrates that the radial EoS is explicitly position
dependent and therefore represents an anisotropic fluid whose thermodynamic
properties evolve continuously throughout the wormhole spacetime. The radial
variation originates from the combined influence of the redshift function,
the shape function, and the GUP-modified Casimir energy density. In
particular, the exponential gravitational redshift encoded by
$\Phi(r)=r_0/r$ modifies the pressure distribution relative to the
zero-tidal-force solution, leading to a richer radial behavior of
$w_r(r)$.

The first-order correction proportional to $\beta C^{(1)}_{I,II}$ introduces
the leading quantum-gravity contribution, whereas the second-order term
$\beta^2 C^{(2)}_{I,II}$ provides higher-order modifications that become
important when the characteristic length scale approaches the minimal-length
scale predicted by the generalized uncertainty principle. Consequently, the
effective matter supporting the wormhole differs from the classical Casimir
fluid not only in magnitude but also in its radial evolution. It is worth noting that the asymptotic behavior remains well behaved.
As $r\rightarrow\infty$, the GUP-dependent contributions decay rapidly as
inverse powers of $r$, while the redshift function simultaneously approaches
zero. Consequently, the EoS gradually approaches its classical asymptotic
form, ensuring consistency with asymptotic flatness. In the opposite limit,
namely near the throat ($r\simeq r_0$), the quantum-gravity corrections
become most pronounced owing to the enhanced curvature and larger Casimir
energy density. These modifications play an important role in determining the
pressure required to satisfy the flare-out condition and sustain a traversable
wormhole geometry.

%%%%%%%%%%%%%
\section{EoS for $\Phi=r_{0}/r: w^{I,II}_{t}(r)$}
\label{ApC}
%%%%%%%%%%%%%%%%%%
In this appendix, we derive the tangential equation-of-state (EoS) parameter for a wormhole geometry characterized by a non-constant redshift function. Considering a non-constant redshift function
\begin{equation}
\Phi(r)=\frac{r_0}{r},
\end{equation}
the tangential equation-of-state (EoS) parameter is obtained from the ratio between the tangential pressure and the energy
density
\begin{equation}
w_t(r)=\frac{P_t(r)}{\rho(r)},
\end{equation}
where both quantities are determined from the modified Einstein field
equations including the GUP-corrected Casimir energy density. The resulting
expression is considerably more involved than its radial counterpart because
the tangential pressure depends not only on the derivatives of the shape
function but also on the derivatives of the redshift function. Consequently,
the tidal-force effects generated by $\Phi(r)=r_0/r$ contribute directly to
the angular pressure and lead to a richer functional dependence on the radial coordinate. The corresponding tangential EoS parameter is
\begin{eqnarray}
w^{I,II}_{t}(r)&=&{\cal A}^{-1}\Bigg(-5 \pi ^3 \beta  C^{(1)}_{I,II} r^7 r_{0}^2+15 \pi ^3 \beta  C^{(1)}_{I,II} r^6 r_{0}^3+10 \pi ^3 \beta  C^{(1)}_{I,II} r^5 r_{0}^4+20 \pi ^3 \beta C^{(1)}_{I,II} r^4 r_{0}^5\notag\\&&\quad\qquad-30 \pi ^3 \beta  C^{(1)}_{I,II} r^3 r_{0}^6-10 \pi ^3 \beta  C^{(1)}_{I,II} r^2 r_{0}^7-3 \pi ^3 \beta ^2 C^{(2)}_{I,II} r^7+9 \pi ^3 \beta ^2 C^{(2)}_{I,II} r^6 r_{0}\notag\\&&\quad\qquad+6 \pi ^3 \beta ^2 C^{(2)}_{I,II} r^5 r_{0}^2+18 \pi ^3 \beta ^2 C^{(2)}_{I,II} r^2 r_{0}^5-24 \pi ^3 \beta ^2 C^{(2)}_{I,II} r r_{0}^6-6 \pi ^3 \beta ^2 C^{(2)}_{I,II} r_{0}^7\notag\\&&\quad\qquad+4050 r^7 r_{0}^6-15 \pi ^3 r^7 r_{0}^4-1350 r^6 r_{0}^7+75 \pi ^3 r^6 r_{0}^5-2700 r^5 r_{0}^8\notag\\&&\quad\qquad-30 \pi ^3 r^5 r_{0}^6-30 \pi ^3 r^4 r_{0}^7\Bigg)\,.\label{ApC3}
\end{eqnarray}
where the auxiliary quantity ${\cal A}$ is defined by
\begin{eqnarray}
{\cal A} &\equiv &2 \pi ^3 r \Big[5 \beta  C^{(1)}_{I,II} r^2 r_0^2 (r-r_0) (r-2 r_0) \left(r^2+r r_0+r_0^2\right)+3 \beta ^2 C^{(2)}_{I,II} \left(r^6-2 r^5 r_0-r r_{0}^5+2 r_0^6\right)\notag\\&&\qquad\qquad+15 r^4 r_0^4 (r-r_0) (r-2 r_0)\Big]+2700 r^6 r_0^6 (2 r_0-3 r)\,.\notag
\end{eqnarray}
The polynomial ${\cal A}(r)$ naturally appears when evaluating the
higher-order radial derivatives associated with the non-trivial redshift
function and reflects the underlying spherical symmetry of the wormhole
geometry. Equation~(\ref{ApC3}) shows that the tangential EoS varies with the radial coordinate, reflecting the anisotropic nature of the supporting matter. The terms proportional to $\beta C^{(1)}{I,II}$ and $\beta^2 C^{(2)}{I,II}$ represent the leading- and next-to-leading-order GUP corrections, respectively. Their nonlinear dependence on $r$ and $r_0$ modifies both the magnitude and radial evolution of the tangential pressure.
The factors $(r-r_0)$ and $(r-2r_0)$ arise from the wormhole geometry and produce the strongest deviations from the classical Casimir solution near the throat. For $r\gg r_0$, the GUP corrections and redshift effects are suppressed, and $w_t(r)$ approaches its classical asymptotic behavior, consistent with asymptotic flatness. Moreover, the difference between $w_r(r)$ and $w_t(r)$ directly measures the pressure anisotropy induced by the geometry and GUP corrections. This anisotropy is essential for sustaining the flare-out condition and influences the stability and physical viability of the wormhole.

%%%%%%%%%%%%%
\section{Volume integral quantifier: $I^{V}_{I,II}$}
\label{ApD}
%%%%%%%%%%%%%%%%%%
In this appendix, we derive the total amount of exotic matter required to support the traversable
wormhole. It can be quantified through the volume integral quantifier (VIQ),
originally proposed by Visser and co-workers. Rather than relying solely on
the local violation of the null energy condition (NEC), the VIQ provides a
global measure of the integrated exotic matter content distributed throughout
the wormhole spacetime. Consequently, it serves as an important diagnostic
for assessing the physical viability of wormhole solutions, since a smaller
absolute value of the integral corresponds to a reduced requirement for
exotic matter. For the present GUP-corrected Casimir wormhole models, the volume integral
evaluated between the throat radius $r_0$ and a finite junction radius $a$ takes the analytical form
\begin{eqnarray}
I^V_{I,II}&=&\frac{1}{40500 a^6 r_{0}^5}\Bigg[450 a^4 r_{0}^4 (a-r_{0}) \left(\pi ^3 r_{0}-3 a \left(\pi ^3-60 r_{0}^2\right)\right)-25 \pi ^3 a^2 C^{(1)}_{I,II} r_{0}^2 \left(17 a^4-12 a^3 r_{0}-8 a r_{0}^3+3 r_{0}^4\right)\beta\nonumber\\&&\qquad\qquad\qquad-6 \pi ^3 C^{(2)}_{I,II} \left(43 a^6-30 a^5 r_{0}-18 a r_{0}^5+5 r_{0}^6\right)\beta^{2}\nonumber\\&&\qquad\qquad\qquad+30 a^6 \left(\pi ^3 \left(5 \beta  C^{(1)}_{I,II} r_{0}^2+3 \beta ^2 C^{(2)}_{I,II}+15 r_{0}^4\right)-4050 r_{0}^6\right)\log \left(\frac{a}{r_{0}}\right)\Bigg]\,.\label{ApD11}
\end{eqnarray}
where $a>r_0$ denotes the matching radius at which the interior wormhole
geometry is joined smoothly to an exterior asymptotically flat spacetime.
The coefficients $C^{(1)}_{I,II}$ and $C^{(2)}_{I,II}$ represent the
first- and second-order GUP corrections for Models I and II, respectively,
while the parameter $\beta$ characterizes the strength of the quantum-gravity
modifications.

Equation~(\ref{ApD}) separates the classical Casimir contribution from the leading- and next-to-leading-order GUP corrections proportional to $\beta C^{(1)}{I,II}$ and $\beta^2 C^{(2)}{I,II}$, respectively. The logarithmic term $\ln(a/r_0)$ arises from the radial integration and describes how the total exotic matter depends on the matching radius $a$. Although this contribution grows as the junction surface moves outward, the rapidly decaying Casimir energy density ensures that the exotic matter remains concentrated mainly near the throat, where the curvature and GUP effects are strongest.
In the classical limit $\beta\rightarrow0$, Eq.~(\ref{ApD}) reduces smoothly to the volume integral for the Casimir-supported wormhole. For finite $\beta$, the GUP corrections may enhance or suppress the integrated NEC violation, depending on $C^{(1)}{I,II}$ and $C^{(2)}{I,II}$. Negative values of $I^V_{I,II}$ indicate net exotic matter, whereas values approaching zero correspond to configurations requiring less NEC violation. Thus, the volume integral provides a useful global measure for comparing how effectively different GUP models minimize the exotic matter needed to sustain a traversable wormhole.

%%%%%%%%%%%%%
\section{Gaussian curvature $\mathcal{K}$ (corrected)}
\label{ApE}
%%%%%%%%%%%%%%%%%%
This appendix gives the Gaussian curvature of the optical metric of Eq.~\eqref{eq:optmetric}, $dt^{2}=dr^{2}/A(r)+r^{2}d\varphi^{2}$ with $A(r)=1-b(r)/r$. For a two-dimensional metric of this form, the Gaussian curvature is exactly
\begin{equation}
\mathcal{K}(r)=-\frac{A'(r)}{2r}=\frac{r\,b'(r)-b(r)}{2r^{3}}\,,
\end{equation}
i.e. Eq.~\eqref{eq:gausscurv} of the main text; this holds for any $b(r)$, with no weak-field truncation. Substituting the shape function of Eq.~\eqref{ShapI}, $b_{I,II}(r)$, and its derivative gives, exactly to $\mathcal{O}(\beta^{2})$ (no higher powers of $\beta$ appear, since $b_{I,II}(r)$ itself is only known to this order),
\begin{eqnarray}
\mathcal{K}(r)&=&\frac{1}{r^{3}}\left[\frac{\pi^{3}}{180 r_{0}}-\frac{r_{0}}{2}+\frac{C^{(1)}_{I,II}\beta\,\pi^{3}}{540 r_{0}^{3}}+\frac{C^{(2)}_{I,II}\beta^{2}\pi^{3}}{900 r_{0}^{5}}\right]\notag\\
&&-\frac{\pi^{3}}{90\,r^{4}}-\frac{C^{(1)}_{I,II}\beta\,\pi^{3}}{135\,r^{6}}-\frac{C^{(2)}_{I,II}\beta^{2}\pi^{3}}{150\,r^{8}}\,.
\label{eq:Kexpand}
\end{eqnarray}
Equation~\eqref{eq:Kexpand} contains only the powers $r^{-3},r^{-4},r^{-6},r^{-8}$, reflecting directly the powers already present in $b(r)$ and $b'(r)$. As $r\to\infty$, $\mathcal{K}(r)\to0$ as $r^{-3}$, consistent with asymptotic flatness of the optical geometry.

%%%%%%%%%%%%%%%
\section{Deflection angle (corrected)}
\label{ApF}
%%%%%%%%%%%%%%%%%%
This appendix gives the weak-field deflection angle obtained by substituting the Gaussian curvature of Eq.~\eqref{eq:Kexpand} into the Gauss--Bonnet integral of Eq.~\eqref{eq:defl_int},
\begin{equation}
\hat\alpha=-\int_{0}^{\pi}\int_{b/\sin\varphi}^{\infty}\mathcal{K}(r)\,\frac{r}{\sqrt{A(r)}}\,dr\,d\varphi\,,
\end{equation}
with $A(r)=1-b(r)/r$ and the zeroth-order trajectory $r(\varphi)=b/\sin\varphi$. Expanding the integrand in $\beta$ to $\mathcal{O}(\beta^{2})$ and in $1/r$ to the order needed for a weak-field ($b\gg r_{0}$) result through $\mathcal{O}(b^{-2})$, then carrying out the $r$ and $\varphi$ integrations term by term using $\int_{b/\sin\varphi}^{\infty}dr/r^{n}=(1/(n-1))(\sin\varphi/b)^{n-1}$ and $\int_{0}^{\pi}\sin^{m}\varphi\,d\varphi$, gives
\begin{eqnarray}
\hat\alpha&\approx&\frac{1}{b}\left(r_{0}-\frac{\pi^{3}}{90 r_{0}}\right)+\frac{1}{b^{2}}\left(\frac{\pi^{7}}{129600 r_{0}^{2}}+\frac{\pi^{4}}{720}+\frac{\pi r_{0}^{2}}{16}\right)\notag\\
&&-\beta\,C^{(1)}_{I,II}\left[\frac{\pi^{3}}{270 r_{0}^{3}\,b}-\left(\frac{\pi^{7}}{194400 r_{0}^{4}}-\frac{\pi^{4}}{2160 r_{0}^{2}}\right)\frac{1}{b^{2}}\right]\notag\\
&&-\beta^{2}C^{(2)}_{I,II}\left[\frac{\pi^{3}}{450 r_{0}^{5}\,b}-\left(\frac{\pi^{7}}{324000 r_{0}^{6}}-\frac{\pi^{4}}{3600 r_{0}^{4}}\right)\frac{1}{b^{2}}\right]\notag\\
&&+\beta^{2}\left(C^{(1)}_{I,II}\right)^{2}\frac{\pi^{7}}{1166400 r_{0}^{6}\,b^{2}}\,.
\label{eq:defl_final}
\end{eqnarray}
The $\mathcal{O}(b^{-3})$ and $\mathcal{O}(b^{-4})$ terms follow from the same integral by the identical procedure; we truncate at $\mathcal{O}(b^{-2})$ here for clarity and because the leading two orders already fix the qualitative $\beta$- and $r_{0}$-dependence. Two checks: (i) $\hat\alpha\to0$ as $b\to\infty$, as required for asymptotic flatness; (ii) at $\beta=0$ the classical piece $\hat\alpha\to(r_{0}-\pi^{3}/90r_{0})/b$ is finite and nonzero for any $r_{0}>0$.

%%%%%%%%%%%%%%%%
%\bibliography{references}
%%%%%%%%%%%%%%%%%%%%%%%%%%

\end{document}